\documentclass{aa}
\usepackage{graphicx}
\usepackage{txfonts}
\usepackage{lineno}
\usepackage{amsmath}    
\usepackage{amssymb}    
\usepackage{cases}  
\usepackage{multirow} 
\usepackage{xcolor} 
\usepackage[normalem]{ulem} 

\usepackage{xfrac}
\DeclareRobustCommand*{\drv}{\mathop{}\!\mathrm{d}}
\DeclareRobustCommand*{\kms}{km\,s$^{-1}$}
\DeclareMathOperator{\Tr}{Tr}

\begin{document} 

   \title{Solar wind charge exchange in cometary atmospheres}

   \subtitle{II. Analytical model}

   \author{Cyril Simon Wedlund\inst{1}
\and Etienne Behar\inst{2,3}
\and Esa Kallio\inst{4}
\and Hans Nilsson\inst{2,3}
\and Markku Alho\inst{4}
\and Herbert Gunell\inst{5,6}
\and Dennis Bodewits\inst{7}
\and Arnaud Beth\inst{8}
\and Guillaume Gronoff\inst{9,10}
\and Ronnie Hoekstra\inst{11}
}

   \institute{Department of Physics, University of Oslo, P.O. Box 1048 Blindern, N-0316 Oslo, Norway\\
              \email{cyril.simon.wedlund@gmail.com}
         \and
             Swedish Institute of Space Physics, P.O. Box 812, SE-981 28 Kiruna, Sweden
        \and
             Lule\aa{} University of Technology, Department of Computer Science, Electrical and Space Engineering, Kiruna,  SE-981 28, Sweden
        \and
            Department of Electronics and Nanoengineering, School of Electrical Engineering, Aalto University, P.O. Box 15500, 00076 Aalto, Finland
        \and
            Royal Belgian Institute for Space Aeronomy, Avenue Circulaire 3, B-1180 Brussels, Belgium
        \and
            Department of Physics, Ume\aa{} University, 901 87 Ume\aa{}, Sweden
        \and 
            Physics Department, Auburn University, Auburn, AL 36849, USA
        \and 
            Department of Physics, Imperial College London, Prince Consort Road, London SW7 2AZ, United Kingdom
                \and
            Science directorate, Chemistry \& Dynamics branch, NASA Langley Research Center, Hampton, VA 23666 Virginia, USA
        \and
            SSAI, Hampton, VA 23666 Virginia, USA   
        \and 
            Zernike Institute for Advanced Materials, University of Groningen, Nijenborgh 4, 9747 AG, Groningen, The Netherlands
            }

   \date{\today}

 
  \abstract
   {Solar wind charge-changing reactions are of paramount importance to the physico-chemistry of the atmosphere of a comet because they mass-load the solar wind through an effective conversion of fast, light solar wind ions into slow, heavy cometary ions. The ESA/\emph{Rosetta} mission to comet 67P/Churyumov-Gerasimenko (67P) provided a unique opportunity to study charge-changing processes in situ.}
   {To understand the role of charge-changing reactions in the evolution of the solar wind plasma and to interpret the complex in situ measurements made by \emph{Rosetta}, numerical or analytical models are necessary.}
   {An extended analytical formalism describing solar wind charge-changing processes at comets along solar wind streamlines is presented. It is based on a thorough book-keeping of available charge-changing cross sections of hydrogen and helium particles in a water gas.}
   {After presenting a general 1D solution of charge exchange at comets, we study the theoretical dependence of charge-state distributions of (He$^{2+}$,~He$^+$,~He$^0$) and (H$^{+}$,~H$^0$,~H$^-$) on solar wind parameters at comet 67P. We show that double charge exchange for the He$^{2+}-$H$_2$O system plays an important role below a solar wind bulk speed of $200$~\kms{} , resulting in the production of He energetic neutral atoms, whereas stripping reactions can in general be neglected. Retrievals of outgassing rates and solar wind upstream fluxes from local \emph{Rosetta} measurements deep in the coma are discussed. Solar wind ion temperature effects at $400$~\kms{} solar wind speed are well contained during the \emph{Rosetta} mission. 
   }
   {As the comet approaches perihelion, the model predicts  a sharp decrease of solar wind ion fluxes by almost one order of magnitude at the location of \emph{Rosetta}, forming in effect a solar wind ion cavity. This study is the second part of a series of three on solar wind charge-exchange and ionization processes at comets, with a specific application to comet 67P and the \emph{Rosetta} mission.
   }

   \keywords{ comets: general -- comets: individual: 67P/Churyumov-Gerasimenko -- instrumentation: detectors -- solar wind, methods: analytical -- solar wind: charge-exchange processes -- Methods: analytical}

   \maketitle
%


\section{Introduction}
Over the past decades, evidence of charge-exchange (CX) reactions has been discovered in astrophysics environments, from cometary and planetary atmospheres to the heliosphere and to supernovae environments \citep{Dennerl2010}. They consist of the transfer of one or several electrons from the outer shells of neutral atoms or molecules, denoted M, to an impinging ion, noted X$^{i+}$, where $i$ is the initial charge number of species X. Electron \emph{\textup{capture}} of $q$ electrons takes the form

\begin{align}
        \textnormal{X}^{i+} + \textnormal{M} &\longrightarrow \textnormal{X}^{(i-q)+} + [\textnormal{M}]^{q+}. \label{eq:capture}
\end{align}

From the point of view of the impinging ion, a reverse charge-changing process is the electron loss (or  stripping\textup{}); starting from species $\textnormal{X}^{(i-q)+}$, it results in the emission of $q$ electrons:

\begin{align}
        \textnormal{X}^{(i-q)+} + \textnormal{M} &\longrightarrow \textnormal{X}^{i+} + [\textnormal{M}]\ +\ qe^- \label{eq:loss}.
\end{align}

For $q=1$, the processes are referred to as one-electron charge-changing reaction; for $q=2$, two-electron or double charge-changing reactions, and so on. The qualifier "charge-changing" encompasses both capture and stripping reactions, whereas "charge exchange" denotes electron capture reactions only.
Moreover, "[M]" refers here to the possibility for compound M to undergo, in the process, dissociation, excitation, and ionization, or a combination of these processes.

Charge exchange was initially studied as a diagnostic for man-made plasmas \citep[][]{Isler1977,Hoekstra1998}. The discovery by \cite{Lisse1996} of X-ray emissions at comet Hyakutake C/1996 B2 was attributed by \cite{Cravens1997} to charge-transfer reactions between highly charged solar wind oxygen ions and the cometary neutral atmosphere. Since this first discovery, cometary CX emission has successfully been used to remotely $(i)$ measure the speed of the solar wind \citep{Bodewits2004ApJ}, $(ii)$ measure its composition \citep{Kharchenko2003}, and thus the source region of the solar wind \citep{Bodewits2007AA,Schwadron2000}, $(iii)$ map plasma interaction structures \citep{Wegmann2005}, and more recently, $(iv)$ to determine the bulk composition of cometary atmospheres  \citep{Mullen2017}.

Observations of charge-exchanged helium, carbon, and oxygen ions were made during the Giotto mission flyby to comet 1P/Halley and were reported by \cite{Fuselier1991}, who used a simplified continuity equation \citep[as in][]{Ip1989} to describe CX processes. \cite{Bodewits2004ApJ} reinterpreted their results with a new set of cross sections. More recently, the European Space Agency (ESA) \emph{Rosetta} mission to comet 67P/Churyumov-Gerasimenko (67P) between August 2014 and September 2016 has provided a unique opportunity for studying CX processes in situ and for an extended period of time \citep{Nilsson2015,CSW2016}. The observations need to be interpreted with the help of analytical and numerical models.

As the solar wind impinges on a neutral atmosphere, either in expansion (comets) or gravity-bound (planets), charge-transfer collisions effectively result in the replacement of the incoming fast (solar wind) ion by a slow-moving (atmospheric) ion \citep{Dennerl2010}. Through conservation of energy and momentum, mass loading of the solar wind occurs and is responsible for the deflection and slowing down of the solar wind ions upstream of the cometary nucleus \citep[see][for comet 67P]{Behar2016b,Behar2016a}. 
For comet 67P, which has a relatively low outgassing rate, the atmosphere is essentially a mixture of H$_2$O, CO$_2$ , and CO molecules \citep{Hassig2015,Fougere2016}. As a first approximation, we only consider capture and stripping collisions in H$_2$O, because this species represents the bulk of the cometary gas during the \emph{Rosetta} mission, except at large heliocentric distances \citep[above about $3$~astronomical units or AU, see][]{Lauter2018}. These reactions result in the production of energetic neutral atoms (ENAs, such as H and~He), which continue to travel in straight lines from their production region, and further interact with the ion and neutral environment.

At comet 67P, evidence of solar wind charge transfer is readily seen in the observations of the Rosetta Plasma Consortium (RPC) ion and electron spectrometers. \cite{Nilsson2015,Nilsson2015AA} and \cite{CSW2016} have reported the detection of He$^+$ ions with the RPC Ion Composition Analyser \citep[RPC-ICA, ][]{Nilsson2007}, arising from incoming charge-exchanged He$^{2+}$ solar wind ions. 

Numerical and analytical models have been developed to account for the detected ion fluxes. \cite{Khabibrakhmanov1997} developed a 1D hydrodynamic model of CX and photoionization at comet 1P/Halley, concluding that the position of the bow shock shifted outward when taking into account single-electron capture of protons in water. \cite{Ekenback2008} used a magnetohydrodynamics (MHD) model to produce images of hydrogen ENA emissions around a comet similar to comet 1P/Halley at perihelion. \cite{CSW2016} proposed in a recent paper a simple 1D analytical model, using only one electron capture reaction (He$^{2+}\rightarrow$~He$^{+}$) to account for the He$^+$ fluxes that were routinely measured by RPC-ICA on board \emph{Rosetta}. The authors showed that from the local measurement of $\textnormal{He}^{+}/\textnormal{He}^{2+}$ flux ratios in the inner coma, it was possible to infer the total outgassing rate of the comet. Comparison with in situ derived outgassing rates by the Rosetta Orbiter Spectrometer for Ion and Neutral Analysis Comet Pressure Sensor (ROSINA-COPS) \citep{Balsiger2007} showed that with these simple assumptions, month-to-month differences between the RPC-ICA-inferred and ROSINA-measured water outgassing rates remained within a factor $2-3$ \citep{Hansen2016}. In parallel, using a new quasi-neutral hybrid model of the cometary plasma environment, \cite{CSW2017} studied the interplay between ionization processes in the formation of boundaries at comet 67P. They showed that CX plays a major role at large cometocentric distances ($>1000$~km at a heliocentric distance of $1.3$~AU), whereas photoionization and electron ionization (sometimes referred to as "electron impact ionization") is the main source of new cometary ions in the inner coma \citep{Bodewits2016}. This is in agreement with observations of electron densities combined with a more precise ionospheric modeling \citep{Galand2016,Heritier2017,Heritier2018}.

During the \emph{Rosetta} mission and while approaching perihelion, the solar wind experienced increasing angular deflection with respect to the Sun-comet line, defining a so-called "solar wind ion cavity" \citep[][noted SWIC for short]{Behar2017}. This is due to the increased cometary outgassing activity and mass loading during that period of time, spanning April to December 2015. As a result, and except for a few occasional appearances due to \emph{Rosetta} excursions at large cometocentric distances, no He$^{2+}$ and He$^+$ signal could be simultaneously detected in the SWIC \citep{CSW2016,Nilsson2017a,Behar2017}.

\begin{figure}
  \includegraphics[width=\linewidth]{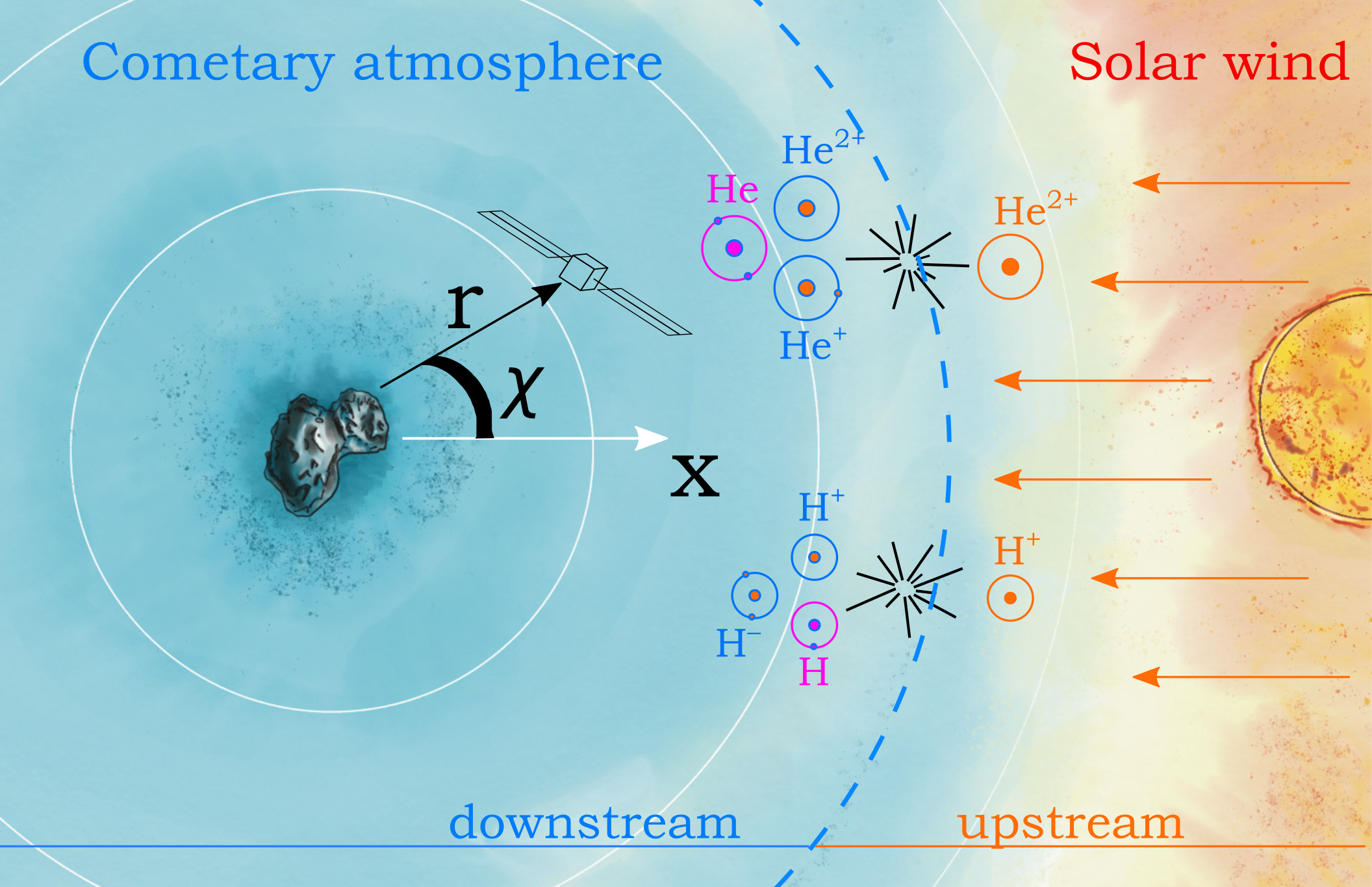}
     \caption{Sketch of Sun-comet CX interactions. The upstream solar wind, composed of H$^{+}$ and He$^{2+}$ ions, experiences CX collisions when impacting the comet's neutral atmosphere, producing a mixture of charged states downstream of the collision. ENAs are depicted in pink. $\chi$ is the solar zenith angle, $r$ the cometocentric distance of a virtual spacecraft, and $x$ points toward the comet-Sun direction, in cometocentric solar equatorial system coordinates. An increasingly deep blue denotes a correspondingly denser atmosphere.
     }
     \label{fig:cometSunsketch}
 \end{figure}

Charge-state distributions and their evolution with respect to outgassing rate and cometocentric distance represent a proxy for the efficiency of charge-changing reactions at a comet such as 67P, as sketched in Fig.~\ref{fig:cometSunsketch}.
In our companion paper \citep[][subsequently referred to as Paper~I]{CSW2018a}, we gave recommended charge-changing and ionization cross sections for helium and hydrogen particles colliding with a water gas.

In this study (referred to as Paper~II), we expand the initial approach expounded in \cite{CSW2016} to include all six main charge-changing cross sections, and present a general analytical solution of the $\text{three}$-component system of helium and hydrogen, with physical implications specific to comets. The forward model expressions are given, and two inversions are proposed, one for deriving the outgassing rate of the comet, one for estimating the upstream solar wind flux from in-situ ion observations. In Section~\ref{sec:discussion}, and using our recommended set of cross sections (see Paper~I), we explore the dependence of the charge-state distribution at comet 67P on heliocentric and cometocentric distances, and solar wind speed and temperature. From geometrical considerations only, we finally make predictions for the charge-state distribution at comet 67P at the location of \emph{Rosetta} (Section~\ref{sec:syntheticCX}).


\section{Solar wind charge distributions at a comet}\label{sec:analytics}

It is well known in the experimental community that charge-state fractions follow a system of coupled differential equations that can be solved analytically: \cite{Allison1958} \citep[and associated erratum][]{Allison1959}, and later \cite{Tawara1973}, for example, give expressions of the charge-state fractions of helium and hydrogen beams in gases for laboratory diagnostic in the measurement of charge-changing cross sections. In these experiments, beams of incoming ions are set to collide, usually in a vacuum chamber, through solid foils or in gases of known characteristics or neutral densities.

In this section, we apply such a formalism to a cometary environment. We give the equations in matrix form for an $(N+1)$-component system of solar wind projectiles, with $N+1$ the number of charge states arising from the charge-changing reactions. We generalize the solution using exponential matrices, and apply this formalism to a three-component charge-changing system between (He$^{2+}$, He$^+$, He$^0$) and (H$^{+}$, H$^0$, H$^-$) solar wind projectiles and a cometary atmosphere, with $N=2$. Inversions of the forward solution include the determination from local observations of the neutral outgassing rate of the comet, as well as that of an estimate of the solar wind upstream flux. Matrices and vectors are denoted in bold font. The nomenclature of the explicit solution is loosely inspired by Allison's, when needed. 

In the following, the CX forward model and its inversions are described for the helium system. For completeness, the solution for the hydrogen system is given in Appendix~\ref{appendix:HydrogenTheory}.

\subsection{General model of charge-changing reactions}

A solar wind plasma species X of initial charge $i$ will undergo electron capture and loss reactions when interacting with cometary neutral species M:
\begin{align}
        \textnormal{X}^{i+}_\textnormal{fast} + \textnormal{M}_\textnormal{slow} &\longrightarrow \textnormal{X}^{(i-q)+}_\textnormal{fast} + [\textnormal{M}]_\textnormal{slow}^{q+}, \\
        \textnormal{X}^{i+}_\textnormal{fast} + \textnormal{M}_\textnormal{slow} &\longrightarrow \textnormal{X}^{(i+q)+}_\textnormal{fast} + [\textnormal{M}]_\textnormal{slow}\ +\ qe^-,
\end{align}
resulting in one reaction in the capture of $q$ electrons by species X and the ionization of neutral compound [M], and in the other, in the loss of $q$ electrons by species X. [M] denotes all possible dissociation, excitation, and ionization stages of species M. In doing so, from the plasma point of view, fast, usually light, solar wind ions are depleted in favor of the production of slow-moving heavy cometary ions because the neutral gas has velocities of about $1$~\kms{} \citep{Hansen2016}, which are added to the solar wind flow. This is one of the basic aspects of solar-wind mass loading \citep{Behar2016a}.

\subsubsection{Continuity matrix system}
In the fluid approximation, the continuity equation for solar wind species X$^{i+}$ of density $n_i$ along bulk velocity $\vec{U}_i$ can be written as
\begin{align}
    \frac{\partial n_i}{\partial t} + \vec{\nabla} \cdot n_i \vec{U}_i = \mathcal{S}_i - \mathcal{L}_i,
\end{align}
with $\mathcal{S}_i$ and $\mathcal{L}_i$ its source and loss terms. To simplify this equation, two assumptions can be made: $(i)$ the upstream solar wind is not time dependent, and so $\partial n_i/\partial t = 0$ (stationary case), and $(ii)$ we assume that all particles of solar wind origin are moving along the solar wind bulk velocity $\vec{U}_i$, with abscissa $s$ in the Sun-comet direction, with no deviation to their initial direction. 
Remarking that particle flux $\vec{F}_i = n_i\vec{U}_i$, we obtain
\begin{align}
    \frac{\drv F_i(s,T_i)}{\drv s} = \mathcal{S}_i(s,T_i) - \mathcal{L}_i(s,T_i),
    \label{eq:continuityEq}
\end{align}
where $T_i$ is the ion temperature of ion species of charge $i$. 

Source and loss terms generally depend on the path $\drv s = U_i\drv t$ that the solar wind ions are having as a bulk (following bulk velocity $\vec{U}_i$ along streamline $s$), but also on the ion temperature $T_i$, that is, the path of the individual ion ($\varv_i\drv t$). We show below that the effect of the temperature of the solar wind ions can be taken into account \textup{\emph{a posteriori}}, using for example Maxwellian-averaged cross sections at a given ion temperature (see Paper~I), in order to mimic the change in efficiency of the reactions. Here, we subsequently assume for simplicity that all ions of different charge have the same temperature, and that $T_i = 0$. Moreover, we have implicitly assumed that all charge states follow the same path; rigorously, charged species will follow, depending on their mass-to-charge ratio, a cycloidal motion driven by the solar wind electromagnetic field, whereas neutral species paths will be unaffected. For simplicity, we assume in the following that all charge states of a solar wind species move with the same bulk velocity (i.e., along solar wind streamlines). This assumption may introduce errors for example in the outgassing rate retrievals presented in Section~\ref{sec:inversion}. In Paper~III, our outgassing rate estimates from ion spectrometer data match those from neutral measurements within a factor $2$, implying a posteriori that to a first approximation, this assumption may hold.

For an initial system of $N+1$ coupled plasma species in different charge states (e.g., the $\text{three}$-component charge system of helium with $N= 2$, He$^{2+}$, He$^{+}$ , and He$^{0}$, or the multiple charge system of oxygen with $N=7$, O$^{7+}$, O$^{6+}$, O$^{5+}$, etc.), source and loss functions for species of charge $i$ can be rewritten as
\begin{align}
    \mathcal{S}_i(s) & = \sum_{j\neq i}^{N} \sigma_{j,i}\, F_j(s)\ n_n(s)\\
    \mathcal{L}_i(s) & = \sum_{j\neq i}^{N} \sigma_{i,j}\, F_i(s)\ n_n(s),
\end{align}
where $n_n(s)$ is the cometary neutral density at coordinate $s$, $\sigma_{j,i}$ is the charge-changing cross section for processes creating a particle of charge $i$, from a corresponding particle of charge $j$ impacting a neutral species: for example, particle He$^{2+}$ ($i=2$) is created from particle He$^+$ ($j=1$) through single electron loss. Similarly, $\sigma_{i,j}$ is the charge-changing cross section representing the main loss from species of charge $i$ to species of charge $j$: for example, particle He$^{+}$ ($i=1$) is undergoing capture of one electron, creating particle He$^{0}$ ($j=0$). The sums defining the source and loss terms for ions in charge state $i$ are performed over all other charge states $j$ (with $j\,\neq\,i$).

Posing that the column density element is $\drv \eta = n_n(s) \drv s$ and dropping the $(s)$ dependence of the variables for convenience, equation~(\ref{eq:continuityEq}) becomes
\begin{align}
    \frac{\drv F_i}{\drv \eta} &= \sum_{j\neq i}^{N+1} \sigma_{j,i}\, F_j - \sum_{j\neq i}^{N+1} \sigma_{i,j}\, F_i.
    \label{eq:continuityEqs}
\end{align}

Assuming that the system is closed and the initial "undisturbed" solar wind flux $F^\textnormal{sw}$ of species X is conserved in a streamline cylinder, the sum of all charge states must remain equal to it:
\begin{align}
    F^\textnormal{sw} = \sum_{j}^{N+1} F_j.
\end{align}

If we express the lowest charge state $l$, in this case, the $(N\,+\,1)^\textnormal{th}$ state, of the initial system of coupled charged species X as the sum of the other charge states, that is, $F_l = F^\textnormal{sw} - \sum_{j\neq l}^N F_j$,
the initial system can then be rearranged and reduced to $N$ coupled equations with $N$ unknowns in matrix form, starting from the highest (fixed) charge state $k$:
\begin{align}
    \frac{\drv\vec{F}(\eta)}{\drv\eta} &= \vec{A}\vec{F}(\eta) + \vec{B}, \label{eq:matrixsysGeneral}
\end{align}
where $\vec{F}$ and $\vec{B}$ are vectors of length $N$, and $\vec{A}$ is an $N\times N$ matrix:
\begin{align}    
    \vec{A} &= \begin{bmatrix} a_{k,k} & a_{k,k-1} & \dots \\ a_{k-1,k} & a_{k-1,k-1} & \dots \\ \dots & \dots & \dots \\ a_{k-N+1,k} & \dots & a_{k-N+1,k-N+1}\end{bmatrix},
    \quad\textnormal{and}\quad \nonumber\\
    \vec{B} &= F^\textnormal{sw}\ \begin{bmatrix} \sigma_{k-N,k} \\ \sigma_{k-N,k-1} \\ \dots \\ \sigma_{k-N,k-N+1} \end{bmatrix}. \nonumber
\end{align}

Charge states $(i,j)$ are here organized as row/column elements $a_{i,j}$ of matrix $\vec{A}$, in order to keep the generality on the charge-state indices. Vector $\vec{B}$ contains the initial condition of the system, with the rate of production of each considered state from the lowest $(N+1)^\textnormal{th}$ state. Posing that the total charge-changing cross section (loss term) of charge state $i$ is
\begin{align}
    \sigma_i &= \sum_{j\neq i}^{N} \sigma_{i,j},  \label{eq:sumXsections}
\end{align}
we can express the diagonal and non-diagonal terms of \vec{A}:
\begin{align}
    a_{i,i} &= -\left(\sigma_{i} + \sigma_{k-N+1,i}\right) \qquad\textnormal{and}\\
    a_{i,j} &= \sigma_{j,i} - \sigma_{k-N,i} \qquad\forall (i\neq j) \in [\textnormal{$N$ charge states}],
\end{align}
for $k$ the (fixed) highest charge state.

We note that the charge-state distributions depend only on the quantity of atmosphere traversed, and thus do not necessarily imply a rectilinear trajectory along the Sun-comet line for the impacting ions.

However, when interpreting our results in Section~\ref{sec:discussion}, the path of the solar wind ions is usually assumed rectilinear along the Sun-comet plane, in the cometocentric solar equatorial system (CSEq) coordinate system \citep[see, e.g.,][]{Glassmeier2017}. In that case, the model is valid for off-$x_\textnormal{CSEq}$-axis solar wind trajectories (as sketched in Fig.~\ref{fig:cometSunsketch}).

\subsubsection{Matrix solution}

The solution of such a system is the sum of the particular solution to the nonhomogeneous system and of the complementary solution to the homogeneous system (assuming $\vec{B} = 0$).
For $\drv\vec{F}/\drv\eta \rightarrow 0$, system~(\ref{eq:matrixsysGeneral}) simply becomes $\vec{A}\vec{F}^\infty + \vec{B} = 0$, where $\vec{F}^\infty$ is none other than the charge distribution at equilibrium (sources and losses in equilibrium), when the solar wind has encountered enough collisions so as to no longer change in charge composition (collisional thickness close to $1$) \citep[see][for laboratory experiments]{Allison1958}. 
In cometary atmospheres, this equilibrium can be reached in practice for high outgassing rates and deep into the coma.
Because matrix $\vec{A}$ is nonsingular (its determinant is non-zero because all charge-changing cross sections are different) and is thus invertible, $\vec{F}^\infty = -\vec{A}^{-1} \vec{B}$ is a particular solution of system~(\ref{eq:matrixsysGeneral}), which now becomes
\begin{equation}
                \frac{\drv\vec{Y}(\eta)}{\drv\eta} = \vec{A} \vec{Y}(\eta),\quad \textnormal{with}\quad \vec{Y}(\eta) = \vec{F}(\eta) - \vec{F}^\infty.
        \label{eq:newmatrixsys}
\end{equation}
The complementary solution to equation~(\ref{eq:newmatrixsys}) with the initial condition $\vec{Y}(0) = \vec{F}(0) - \vec{F}^\infty$ is $\vec{Y}(\eta) = e^{\vec{A}\eta}\vec{Y}(0)$, with matrix exponential $e^{\vec{A}\eta}~\hat{=}~\sum_{k=0}^{\infty}~(\vec{A}\eta)^{k}/k!$ a fundamental matrix of the system. 

Finally, the solution for the charge distribution column vector \vec{F} function of the column density $\eta$ is
\begin{align}
                \vec{F}(\eta) &= \vec{F}^\infty + e^{\vec{A}\eta}\, \left(\vec{F}(0) - \vec{F}^\infty\right),\label{eq:expomatrixsolution}\\
    \textnormal{with:} \quad
    \vec{F}^\infty &= -\vec{A}^{-1} \vec{B}.\nonumber
\end{align}

The matrix exponential can be calculated by $e^{\vec{A}\eta}~=~\vec{S}e^{\vec{\Lambda}\eta}\vec{S}^{-1}$, where $\vec{\Lambda}$ is the diagonal matrix of the homogeneous system (whose diagonal elements are the eigenvalues), and $\vec{S}$ is the matrix of passage (whose columns are the eigenvectors), so that $\vec{A} = \vec{S}\vec{\Lambda}\vec{S}^{-1}$.

Result~(\ref{eq:expomatrixsolution}) is valid for any system of charged species, with different charge states arising from charge-changing reactions (electron capture and loss) with the neutral atmosphere of an astrophysical body such as a comet or planet. 
This model may include the calculation of the fluxes for high-charged states of atoms in the solar wind, such as oxygen (O$^{7+}$, O$^{6+}$, $\dots$), and carbon (C$^{6+}$, C$^{5+}$, $\dots$), responsible for X-ray emissions at comets and planets \citep{Cravens2009}.

This solution is also applicable to simpler charge-changing systems such as solar wind helium particles (He$^{2+}$,~He$^{+}$,~He$^{0}$), for which we present an explicit solution below. For completeness, the similar solution for the hydrogen (H$^{+}$,~H$^{0}$,~H$^{-}$) system is also given in Appendix~\ref{appendix:HydrogenTheory}.

\subsection{Application to the helium system} \label{sec:Helium}
As previously, let projectile species be numbered by their charge, so that He$^{2+}$, He$^+$ and He$^0$ have $2$, $1$, and $0$ charges, respectively. Through a combination of limited column densities upstream of the comet, expectedly small cross sections, and reduced species lifetimes against autodetachment, lower charge states of helium \citep[such as the short-lived excited state of the He$^-$ anion, see][]{Schmidt2012} are neglected. 

For the $\text{three}$ charge states of helium, the six relevant cross sections $\sigma_{ij}$, here with $i$ and $j$ the starting and end charges, are
\begin{align*}
        \sigma_{21}:&\ \textnormal{He}^{2+} \longrightarrow \textnormal{He}^+ &\ \textnormal{single capture} \\
    \sigma_{20}:&\ \textnormal{He}^{2+} \longrightarrow \textnormal{He}^0 &\ \textnormal{double capture}\\    
    \sigma_{12}:&\ \textnormal{He}^{+} \longrightarrow \textnormal{He}^{2+} &\ \textnormal{single stripping} \\
    \sigma_{10}:&\ \textnormal{He}^{+} \longrightarrow \textnormal{He}^0 &\ \textnormal{single capture} \\
    \sigma_{02}:&\ \textnormal{He}^{0} \longrightarrow \textnormal{He}^{2+} &\ \textnormal{double stripping}\\
    \sigma_{01}:&\ \textnormal{He}^{0} \longrightarrow \textnormal{He}^{+} &\ \textnormal{single stripping}
\end{align*}

We define for each charge state the total charge-changing cross sections, as in equation~(\ref{eq:sumXsections}):
\begin{equation*}
        \begin{split}
                \sigma_2 &= \sigma_{21} + \sigma_{20} \quad \textnormal{for He}^{2+}\\ 
        \sigma_1  &= \sigma_{12} + \sigma_{10} \quad \textnormal{for He}^{+}\\
        \sigma_{0} &= \sigma_{02} + \sigma_{01} \quad \textnormal{for He}^{0}\\
        \sum{\sigma_{ij}} &= \sigma_2 + \sigma_1 + \sigma_{0},
        \end{split}
\end{equation*}
with $\sum{\sigma_{ij}}$ the sum of all six cross sections.

\subsubsection{Matrix system}
With these notations, for $N=2$ and with respect to column density $\eta$, matrix system~(\ref{eq:matrixsysGeneral}) becomes
\begin{align}
                \frac{\drv\vec{F}(\eta)}{\drv\eta} = \vec{A}\vec{F}(\eta) + \vec{B}, \label{eq:HeliumMatrixSys}\\
                \textnormal{with}\quad \begin{cases}
                \vec{A} = \begin{bmatrix} a_{22} & a_{21}\\ a_{12} & a_{11} \end{bmatrix} \\
        \vec{B} = F^\textnormal{sw} \begin{bmatrix} \sigma_{02} \\ \sigma_{01} \end{bmatrix}
        \end{cases}
        \textnormal{and}\quad \vec{F}(0) = \begin{bmatrix} F^\textnormal{sw} \\ 0 \end{bmatrix}.\nonumber
\end{align}

The matrix elements $a_{ij}$ are, dropping the commas for clarity:

 \begin{equation*}
        \begin{split} 
                a_{22} &= -(\sigma_2 + \sigma_{02}), &\qquad a_{21} &= \sigma_{12} - \sigma_{02}, \\
        a_{12} &= \sigma_{21} - \sigma_{01}, &\qquad  a_{11} &= -(\sigma_1 + \sigma_{01}).
        \end{split}
 \end{equation*}
 
In these new notations, we remark also that $\sum{\sigma_{ij}}~=~-(a_{22}~+~a_{11})$.

Fluxes $\vec{F}$ will depend on the initial charge distribution of the incoming undisturbed solar wind. Far upstream of the cometary nucleus ($\eta = 0$), the solar wind is assumed to be composed in this case of He$^{2+}$ ions only, so that $\vec{F}(0)~=~\left[ \begin{smallmatrix} F_2(0) & F_1(0) \end{smallmatrix} \right]^\top~=~\left[ \begin{smallmatrix} F^\textnormal{sw} & 0 \end{smallmatrix} \right]^\top$. A similar assumption can be made separately with protons. We now normalize our local fluxes to the initial solar wind flux by setting $F^\textnormal{sw}=1$ in the following: at the end of our calculations, we then simply need to multiply the final fluxes by $F^\textnormal{sw}$ to obtain the non-normalized quantities.

\subsubsection{Explicit solution}

The complementary solution of the homogeneous solution is obtained by solving the eigenvalue equation $\vec{A}\vec{v} = \lambda\vec{v}$, with $\vec{v}$ the eigenvector associated with the eigenvalue $\lambda$. The characteristic polynomial
\begin{align*}
    p(\lambda)~=~\det(\vec{A}-\lambda\vec{I})~=~\lambda^2~-~\Tr{\vec{A}}\,\lambda~+~\det\vec{A}
\end{align*} 
yields two real eigenvalues \citep{Allison1958}, which are
\begin{align}
\lambda_\pm~=~\frac{1}{2}(a_{11}+a_{22})\pm q = -\frac{1}{2}\sum{\sigma_{ij}} \pm q,    
\end{align}
when posing $q=\frac{1}{2}\sqrt{(a_{22}-a_{11})^2 + 4\,a_{12}a_{21}}$.  

Matrix $\vec{A}$ can then be eigen-decomposed into $\vec{A} = \vec{S}\vec{\Lambda}\vec{S}^{-1}$: 
\begin{equation*}
        \begin{split}
                \vec{\Lambda} &= \begin{bmatrix} \lambda_{-} & 0\\ 0 &  \lambda_{+} \end{bmatrix},\\
        \vec{S} &= \frac{1}{a_{12}}\begin{bmatrix} t - q & t + q\\ a_{12} & a_{12} \end{bmatrix},\\
                \vec{S}^{-1} &= \frac{1}{2q}\begin{bmatrix}  -a_{12} & t+q\\ a_{12}  & -t+q \end{bmatrix},\\
        \textnormal{with}\quad t = \frac{1}{2}(a_{22}-a_{11}).
    \end{split}
\end{equation*}
The matrix exponential, expressed with the use of hyperbolic sine functions, is finally
\begin{equation*}
        \begin{split}
                e^{\vec{A\eta}} &= \vec{S}e^{\vec{\Lambda}\eta}\vec{S}^{-1} \\
        &= \frac{1}{q}\begin{bmatrix} t\sinh{(q\eta)} + q \cosh{(q\eta)} & a_{21}\sinh{(q\eta)}\\ a_{12}\sinh{(q\eta)} & -t \sinh{(q\eta)} + q\cosh{(q\eta)}  \end{bmatrix} \\
        &\quad\times e^{-\frac{1}{2}\sum{\sigma_{ij}}\,\eta}.
    \end{split}
\end{equation*}

Extended to the charge fraction $\text{three}$-component column vector $\vec{F} = \left[ \begin{smallmatrix} F_2 & F_1 & F_0 \end{smallmatrix} \right]^\top$, the solution of system~(\ref{eq:HeliumMatrixSys}), a combination of exponential functions, can then be written in the following final form \citep[equivalent to that of][for a normalized ion beam]{Allison1958}:
\begin{align}
    \vec{F} &= F^\textnormal{sw} \left(\vec{F}^{\infty} + \frac{1}{2q}\left( \vec{P}\ e^{q \eta} - \vec{N}\ e^{-q \eta} \right)\ e^{-\frac{1}{2} \sum{\sigma_{ij}}\,\eta}\right) \label{eq:HeliumSolution} \\\textnormal{with}\nonumber \\
         \vec{F}^\infty &= \begin{bmatrix} F_2^\infty \\ F_1^\infty \\ F_0^\infty \end{bmatrix} = \begin{bmatrix} \multirow{2}{*}{$-\vec{A}^{-1} \vec{B}$} \\ \\ 1-\sum_{i\neq0} F_i^\infty\end{bmatrix} \nonumber\\
         &= \frac{1}{D}\begin{bmatrix}  -a_{11}\sigma_{02}+a_{21}\sigma_{01} \\a_{12}\sigma_{02} - a_{22}\sigma_{01}\\ \sigma_{02}(a_{11}-a_{12}) + a_{22}(a_{11}+\sigma_{01}) - a_{21}(a_{12} + \sigma_{01})\end{bmatrix},\nonumber\\
  \vec{P} &= \begin{bmatrix} P_2 \\ P_1 \\ P_0 \end{bmatrix} = \begin{bmatrix} (t + q)\left(1 - F_2^\infty\right) - a_{21} F_1^\infty\\ a_{12}\left(1 - F_2^\infty\right) + (t - q) F_1^\infty \\ -(t+q+a_{12})\left(1 - F_2^\infty\right) - (t-q-a_{21}) F_1^\infty  \end{bmatrix},\nonumber\\
  \vec{N} &= \begin{bmatrix} N_2 \\N_1 \\ N_0 \end{bmatrix} = \begin{bmatrix} (t - q)\left(1 - F_2^\infty\right) - a_{21} F_1^\infty\\ a_{12}\left(1 - F_2^\infty\right) + (t + q) F_1^\infty\\ -(t-q+a_{12})\left(1 - F_2^\infty\right) - (t+q-a_{21}) F_1^\infty  \end{bmatrix},\nonumber\\
  \textnormal{recalling}\nonumber \\
  t &= \frac{1}{2} (a_{22}-a_{11}), \quad q =\frac{1}{2}\sqrt{(a_{22}-a_{11})^2 + 4\,a_{12}a_{21}}\quad \textnormal{and}\nonumber\\
  \sum{\sigma_{ij}} &= -(a_{22} + a_{11}),\nonumber\\
  \textnormal{and}\nonumber\\
  \vec{A}^{-1} &= \frac{1}{D} \begin{bmatrix} a_{11} & -a_{21}\\ -a_{12} & a_{22} \end{bmatrix}, \quad\textnormal{where}\nonumber\\
  D &= \det{\vec{A}} = a_{11}a_{22}-a_{12}a_{21}.\nonumber
\end{align}

The equilibrium flux $\vec{F}^\infty = -\vec{A}^{-1}\vec{B}$ depends only on cross sections and is given here in full for convenience:
\begin{align}
    \vec{F}^\infty = \begin{cases}
        F_2^\infty& = \frac{(\sigma_{12}+\sigma_{10})\,\sigma_{02}\ +\ \sigma_{12}\,\sigma_{01}}{(\sigma_{12}+\sigma_{10}+\sigma_{01})\,(\sigma_{21}+\sigma_{20}+\sigma_{02})\ +\ (\sigma_{02}-\sigma_{12})\,(\sigma_{21}-\sigma_{01})}\\[0.5em]
        F_1^\infty& = \frac{(\sigma_{21}+\sigma_{20})\,\sigma_{01}\ +\ \sigma_{21}\,\sigma_{02}}{(\sigma_{12}+\sigma_{10}+\sigma_{01})\,(\sigma_{21}+\sigma_{20}+\sigma_{02})\ +\ (\sigma_{02}-\sigma_{12})\,(\sigma_{21}-\sigma_{01})}\ .\\[0.5em]
        F_0^\infty& = 1 - F_2^\infty-F_1^\infty
    \end{cases}\label{eq:equilibriumFluxHelium}
\end{align}

In practice, it is convenient to normalize the fluxes to the upstream solar wind flux ($F^\textnormal{sw} = 1$): the calculated charge-state distributions are in this case comprised between $0$ and $1$.

\subsection{System inversion}\label{sec:inversion}

We present here two types of inversions of systems~(\ref{eq:HeliumSolution}) and Appendix~\ref{eq:HydrogenSolution} to retrieve from cometary observations some important information on the neutral outgassing rate \citep[as in][]{CSW2016}, and on the solar wind upstream conditions.

\subsubsection{Outgassing rate}
A first inversion of the helium system~(\ref{eq:HeliumSolution}) or hydrogen system~(\ref{eq:HydrogenSolution}) consists of extracting the water outgassing rate $Q_0$ from the species fluxes measured by an ion/ENA spectrometer immersed into the atmosphere of a comet. The following development is applied to the helium system and the simultaneous detection of He$^{2+}$ and He$^+$ ions (as in RPC-ICA solar wind measurements), but can be easily extended to any species fluxes (e.g., H$^0$/H$^+$ or H$^-$/H$^+$ for hydrogen).

Ideally, a normalized quantity should be used so that the efficiency of the CX is taken into account without reference to the initial solar wind flux. The ratio $F_1/F_2$ fulfills this criterion \citep{CSW2016}. In equation~(\ref{eq:HeliumSolution}), we then set $F^\textnormal{sw} = 1$. 

The number density of neutrals, assuming a spherically symmetric gas expansion at constant speed $\varv_0$ (m~s$^{-1}$) \citep{Haser1957} and a production rate $Q_0$ (s$^{-1}$) of neutrals $n$, is
\begin{equation}
     \begin{split}
                n_n(r) &=\frac{Q_0}{4\pi \varv_0\ r^2}\label{eq:HaserSimple}
        \end{split},
\end{equation}
with $r = \sqrt{x^2+y^2+z^2}$ the cometocentric distance. We have neglected here the usual exponential term to account for the decay of neutrals at large cometocentric distances, $e^{-(r-r_c) k_\textnormal{p}^\textnormal{T}/\varv_0}$, with $k_\textnormal{p}^\textnormal{T}$ the total photodestruction rate of neutral species $n$ \citep{CSW2016}, since it only accounts in the calculation of the column density for less than $2$\% difference at the close cometary distances usually probed by \emph{Rosetta} (i.e., for cometocentric distances within a few tens up to $500$~km or so). More self-consistent approaches \citep{Festou1981,Combi2004}, taking into account the collisional part of the cometary coma, where the neutral gas moves at slower speeds (with parent, dissociated daughter and grand-daughter species having different ejection speeds), give a different column density of neutrals than the Haser-like profile above, with respect to cometocentric distance. 
However, for our demonstration, and given the uncertainties on several collisional parameters, a Haser-like model gives a reasonable first guess of the neutral distribution \citep{Combi2004}.

The outgassing rate appears as a variable in the column density $\eta$. In the simple case of a rectilinear motion of the solar wind ions along the Sun-comet line, the column density depends on the solar zenith angle $\chi$ \citep{Beth2016}:
\begin{equation}
     \begin{split}
                \eta(r,\chi) &= \int_{r\cos\chi}^{+\infty} n_n(s)\, \drv s = \frac{Q_0}{4\pi \varv_0\ r} \frac{\chi}{\sin\chi} \label{eq:columnDensity}\\
                        &= \frac{Q_0}{\varv_0}\ \epsilon(r,\chi)
        \end{split},
\end{equation}
where $\varv_0$ is the average speed of the outgassing neutrals. $\chi$ (in units of radians) is defined in the spherically symmetric case as the angle between the local $+x$ direction on the comet-Sun line and the Sun, so that $\chi~=~\arccos{(x/r)}$. The quantity $\epsilon(r,\chi)$ thus only depends on the geometry of the encounter, with the physics of the gas production contained in outgassing rate $Q_0$ and neutral velocity $\varv_0$. 

With these notations, system~(\ref{eq:HeliumSolution}) can be rearranged as
\begin{align}
                \mathcal{R} = \frac{F_1 \left(1 - F_1^{\infty}/F_1\right) }{F_2 \left(1 -F_2^{\infty}/F_2\right)} = \frac{\left( P_1\ e^{q Q_0\epsilon/\varv_0} - N_1\ e^{-q Q_0\epsilon/\varv_0} \right)}{\left( P_2\ e^{q Q_0\epsilon/\varv_0} - N_2\ e^{-q Q_0\epsilon/\varv_0} \right)},
\end{align}
which is of the form $\mathcal{R} = \left( P_1 y - N_1/y\right)/\left( P_2 y - N_2/y\right)$, posing $y = \exp({q Q_0\epsilon/\varv_0})$. The equation has two roots, for which we only keep the positive one, since the discriminant of the equation is itself always positive: $\Delta = -4\, (N1-\mathcal{R}N_2)/(P1-\mathcal{R}P_2) \geq 0$ is equivalent to $-F_1^\infty /(1-F_2^\infty) \leq \mathcal{R}$, which is always fulfilled.

The solution for $Q_0$ becomes
\begin{align}
                Q_0 = \varv_0 \frac{\ln\left(\frac{N_1-\mathcal{R} N_2}{P_1-\mathcal{R} P_2}\right)}{2q\, \epsilon(r,\chi)}\label{eq:outgassingRetrieved}.
\end{align}

In certain conditions, ratio $\mathcal{R}$ can be simplified to reflect the direct in situ measurements made by an ion spectrometer, whereas avoiding reference to the initial upstream solar wind flux, a piece of information usually out of instrumental reach. Thus, we can remark that
\begin{equation*}
        \mathcal{R} \rightarrow \frac{F_1}{F_2}, \quad\textnormal{when}\quad \frac{F_i^\infty}{F_i} \ll 1,\ \textnormal{for}\ i=1,2.
\end{equation*}

This relation is in practice observed well for solar wind speeds below $400$~\kms{} and for cometocentric distances between $10$~km and $500$~km, which are the typical distances covered  by the spacecraft \emph{Rosetta} while outside of the solar wind ion cavity (SWIC). The exact range of validity of this assumption is discussed later in Section~\ref{sec:syntheticOutgassing}.

Following \cite{CSW2016}, it is interesting to note that when only $\textnormal{He}^{2+}\rightarrow\textnormal{He}^+$ ($2\rightarrow1$) reactions are taken into account (no electron loss or double capture) and He$^0$ atoms are neglected, system~(\ref{eq:HeliumMatrixSys}) is greatly simplified, and leads to the following expression of $Q_0$:
\begin{equation}
     \begin{split}
                Q_0 &= \varv_0 \frac{\ln\left(\mathcal{R} + 1\right)}{\sigma_{21}\, \epsilon(r,\chi)}.\label{eq:outgassingCSW2016}
        \end{split}
\end{equation}

This expression is not self-consistent within the (He$^{2+}$,~He$^+$) system since the loss term from He$^+$ ions is not considered, and leads to an underestimate of the final outgassing rate. In practice, this expression remains useful in order to give a first indication of the cometary outgassing rate \citep{CSW2016}. 

A third expression of $Q_0$, when no electron losses are taken into account, is proposed in Appendix~\ref{appendix:reducedSystem} as a simple compromise between expressions~(\ref{eq:outgassingRetrieved}) and (\ref{eq:outgassingCSW2016}). This is suitable for most of the \emph{Rosetta} mission.

\subsubsection{Upstream solar wind flux}
At comets, the knowledge of the usptream (unperturbed) solar wind conditions when the spacecraft is deeply embedded in the cometary neutral atmosphere can be difficult to estimate. From the systems of equations presented above and a local observation of the ion fluxes, we show that it is possible, however, to retrieve the upstream solar wind flux, assuming no solar wind deceleration \citep[consistent with the observations of][at comet 67P]{Behar2016a} and a spherically symmetric outgassing. For a more precise approach, the trajectories of ions can be calculated using, for example, a hybrid plasma model \citep{CSW2017}.

An initial solar wind composed of $\alpha$ particles or protons will have a flux $\vec{F}(0)=\left[\begin{smallmatrix} F_i^\textnormal{sw} & 0 & 0 \end{smallmatrix} \right]^\top$ ($i = 2$ or $i = 1$ for He$^{2+}$ or H$^+$). Following a local measurement of the flux of $\alpha$ particles or protons $F_i(r_\textnormal{pos})$ made at position $r_\textnormal{pos}$, systems~(\ref{eq:HeliumSolution}) and ~(\ref{eq:HydrogenSolution}) become
\begin{equation}
     \begin{split}
                F_i(r_\textnormal{pos}) &= F_i^{\infty} + \frac{1}{2q}\left( P_i\ e^{q \eta} - N_i\ e^{-q \eta} \right)\ e^{-\frac{1}{2} \sum{\sigma_{ij}}\,\eta}\\
\textnormal{with} \\
            P_i &= (t + q)\left(F_i^\textnormal{sw} - F_i^\infty\right) - a_{i,i-1} F_{i-1}^\infty, \\
            N_i &= (t - q)\left(F_i^\textnormal{sw} - F_i^\infty\right) - a_{i,i-1} F_{i-1}^\infty.
        \end{split}
\end{equation}

Solving for $F_i^\textnormal{sw}$, the solar wind upstream flux is simply
\begin{align}
                F_i^\textnormal{sw} &= \frac{F_i(r_\textnormal{pos})}{F_i^{\infty} + \frac{1}{2q}\left( P_i\ e^{q \eta} - N_i\ e^{-q \eta} \right)\ e^{-\frac{1}{2} \sum{\sigma_{ij}}\,\eta}}.
    \label{eq:inversionSW}
\end{align}

The initial solar wind flux can thus be retrieved with the additional knowledge of the local cometary density, comprised in $\eta$. 

It is also useful to note, as in Sections~\ref{sec:Helium} and Appendix~\ref{appendix:HydrogenTheory}, that this formula is valid for any trajectory of the incoming solar wind ions because it depends only on the column density $\eta$ traversed. However, deep inside the cometary magnetosphere, the solar wind ions are strongly deflected, and owing to the changes in local magnetic field magnitude and direction, the normal cycloidal motion will be highly disturbed. This implies that in the simplistic assumption of a rectilinear motion along the Sun-comet line of the incoming solar wind ions, the retrieved solar wind upstream flux will be underestimated by this method. Self-consistent modeling taking into account all charge-changing reactions, using hybrid \citep{CSW2017} or multi-fluid MHD models \citep{Huang2016}, can overcome this caveat.


\section{Results and discussion}\label{sec:discussion}

Following the analytical expression of the solar wind charge distribution in the case of a comet (Section~\ref{sec:analytics}), paired with the determination of the cross-section sets in water (Paper~I), we now turn to investigating the efficiency of charge-transfer reactions with respect to the solar wind proton and $\alpha$ particles. We do this from the point of view of equilibrium charge fractions, and the variations in two solar wind-cometary parameters: the outgassing rate (depending on heliocentric distance), and the solar wind speed.

In the following, we assume for simplicity a motion of the solar wind along the Sun-comet line, that is, no deflection or slowing-down of the solar wind takes place, and no magnetic pile-up region forms upstream of the nucleus. Consequently, the initial solar wind along the Sun-comet line, containing solely (He$^{2+}$,~H$^+$), becomes a mixture of their three respective charge states. Moreover, unless otherwise stated, we adopt normalized quantities so that fluxes are comprised between $0$ and $1$ and the initial solar wind flux is set to unity, that is, $F^\textnormal{sw}=1$ in the analytical solutions.

\begin{figure*}
  \includegraphics[width=\linewidth]{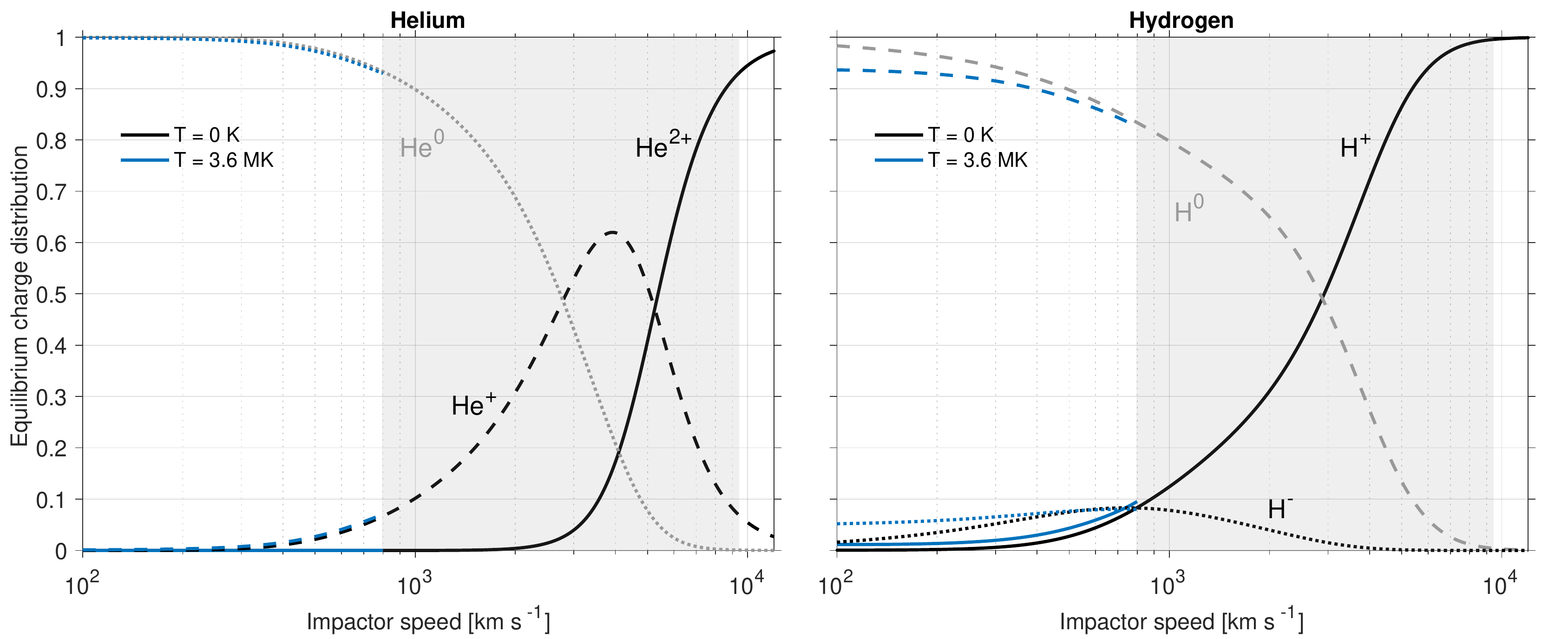}
     \caption{Normalized equilibrium charge fractions of helium (left) and hydrogen (right) in H$_2$O gas as a function of solar wind speed. The fraction of energetic neutral atoms (He$^0$ and H$^0$) is indicated as a gray line. Corresponding Maxwell-averaged solar wind distributions for an ion temperature $T~=~3.6~\times~10^6$~K are plotted in blue in the  $100-800$~\kms{} solar wind speed range. The grayed-out region above $800$~\kms{} shows where the temperature effects are not calculated because of fitting limitations.
     }
     \label{fig:EqChargeFractions}
 \end{figure*}

\subsection{Equilibrium charge-state distribution}\label{sec:EqChargeDistribution}
Figure~\ref{fig:EqChargeFractions} shows the equilibrium charge distributions for helium and for hydrogen, that is, the charge fractions reached at equilibrium in case of the CX mean free path $\sfrac{1}{n_n\sigma_\textnormal{CX}}\,\ll\,1$. As shown in Section~\ref{sec:Helium} and Appendix~\ref{appendix:HydrogenTheory} (equations~(\ref{eq:HeliumSolution}) and (\ref{eq:equilibriumFluxHelium}) for helium, and (\ref{eq:HydrogenSolution}) for hydrogen), these fractions only depend on a linear combination of the cross sections, which themselves vary with impact speed and solar wind ion temperatures; they do not depend on the initial composition of the impacting solar wind. They thus give insight into how efficient the combined charge-changing processes are when energetic hydrogen or helium ions hit a dense atmosphere, or in a controlled environment in laboratory experiments such as charge-equilibrated Faraday cages, where a thin metal foil of thickness $>0.3~\mu$g~cm$^{-2}$ is typically used to achieve equilibrium \citep{Tawara1973}. 

Calculations were performed for monochromatic solar wind beams (i.e., with an equivalent Maxwellian temperature of $0$~K, black and gray lines in Fig.~\ref{fig:EqChargeFractions}), and for a solar wind with a Maxwellian temperature of $3.6\times10^6$~K (thermal velocity $\varv_\textnormal{th}=300$~\kms{} for H$^+$, $150$~\kms{} for He$^{2+}$ ions) that is representative of a typical heating at a bow shock-like structure (blue curves in Fig.~\ref{fig:EqChargeFractions}). The equilibrium charge-state distributions were calculated using the Maxwellian-averaged cross-section fits given in Paper~I that are valid between $100-800$~\kms{} impactor speeds.

In the helium case, He$^{2+}$ ions dominate at very high energies (speeds above $10\,000$~\kms{}) but start to charge-transfer into a mixture of He$^+$ ions and He$^0$ atoms below. He$^+$ ions dominate in a narrow range around $3\,000-5\,000$~\kms{} impact speed where He$^0$ and He$^{2+}$ species make up only $20$\% each of the charge state. In the typical impactor speeds of interest in solar wind-comet studies ($100-800$~\kms{}), the beam is composed almost exclusively of neutral He$^0$ species. The effect of the solar wind temperature is marginal on the charge distributions (blue curves in Fig.~\ref{fig:EqChargeFractions}).

Similarly, in the hydrogen case, H$^+$ ions dominate above $500$~\kms{} impact speed, whereas energetic neutral H$^0$ atoms start to dominate for all speeds below $200$~\kms{}, including in the solar wind speed region. H$^-$ anions make up below $10$\% of the total charge at any energy. In contrast to the helium case, however, the effect of the solar wind temperature on the hydrogen charge distributions becomes quite noticeable, especially below $500$~\kms{}: compared to the monochromatic solution, the H$^{0}$ fraction is $5\%$ lower at $100$~\kms{} solar wind speed, whereas those of H$^-$ and H$^+$ increase by a factor $3$ and $25$ at the same speed (although the proportion of H$^+$ to the total distribution remains very low). This behavior is due to the electron capture and loss cross sections of H$^0$, which peak at high energies, being favored over other reactions, effectively populating H$^+$ and H$^-$ ions (see also Paper~I).
Overall, in both systems, most initial solar wind ions will have charge transferred to their corresponding neutral atom for velocities below $2\,000$~\kms{} by the time they reach equilibrium. 
 
\subsection{Charge distribution at a comet}
The composition of the beam with respect to cometocentric distance in typical cometary and solar wind conditions is explored below.
Atmospheric composition and outgassing rates typical of comet 67P are used throughout. 

\subsubsection{Atmospheric composition}
Charge-exchange reaction effects are cumulative in nature, and as we showed, they depend on the column density of neutrals. A light neutral species such as H or O (arising from photodissociation of H$_2$O, OH, CO$_2$ , or CO), will dominate the coma far upstream of the cometary nucleus; such a minor species in the inner coma may thus play a non-negligible role in the removal of fast solar wind ions a few thousand kilometers upstream of the nucleus because of its large-scale distribution. Above $200\,000$~km for a cometary outgassing rate of $10^{28}$~s$^{-1}$, H may become the major neutral species. This is especially relevant because resonant and semi-resonant reactions, such as $\textnormal{H}^++\textnormal{H}\rightleftharpoons\textnormal{H}+\textnormal{H}^+$, have large electron capture cross sections. 
The resonant one-electron capture cross sections for $\textnormal{H}^+ + \textnormal{X}$ at $U_\textnormal{sw}=450$\,km/s impact speed ($1$\,keV/u) is about the same in X=H or in H$_2$O: $\sigma_{10}(\textnormal{H})\approx19\times10^{-20}$\,m$^2$ \citep{Tawara1985}, compared to
$\sigma_{10}(\textnormal{H}_2\textnormal{O})\approx17\times10^{-20}$\,m$^2$ (see Paper~I). Moreover, because cross sections for resonant processes continue to increase with decreasing energies \citep{Banks1973a}, heating through a bow shock structure is expected to have a relatively small effect on the efficiency of CX reactions such as $\textnormal{H}^++\textnormal{H}$ (see the discussion on Maxwellian-averaged cross sections in Paper~I).

 We now evaluate how much the solar wind proton flux decreases as a result of proton-hydrogen CX. We first use a generalized Haser neutral model such as that of \cite{Festou1981}, taking into account the photodissociated products of water, and apply it to comet 67P at $1.3$~AU ($\sim7\times10^{27}$~s$^{-1}$) for maximum effect. We then calculate the column density of hydrogen along the Sun-comet line up to $10\times10^6$~km. We find that including H and O and calculating the CX encountered by solar wind protons diminishes the expected solar wind flux at $1\,000$~km by $2\%$ with respect to the case where we include H$_2$O only. For lower solar wind speeds ($200$~\kms{}), this decrease remains below $2.5\%$. A similar calculation for He$^{2+}$ ions in He$^{2+}-$H reactions shows that the $\alpha$ particle flux decrease remains below $0.5\%$ at $1$~keV/u impact energy.
 
 These results imply that at comet 67P, the inclusion of the photodissociated products of H$_2$O has a very weak effect on the overall solar wind CX efficiency and the conversion of solar wind protons and $\alpha$ particles into their ENA counterpart. Consequently, the hydrogen and oxygen cometocorona is neglected in the following discussion.
 It is interesting to note, however, that this conclusion may differ between comets (because of different atmospheric composition and activity levels) and between stages of their orbit. Because their outgassing rate is more than two orders of magnitude higher than that of comet 67P at perihelion, comet 1P/Halley and comet C1995~O1/Hale-Bopp have an extended hydrogen corona that does play a non-negligible role at large cometocentric distances \citep{Bodewits2006}. 

During the later part of the \emph{Rosetta} mission, CO$_2$, and to a lesser extent CO, started to dominate the neutral coma over H$_2$O \citep{Fougere2016,Lauter2018}. At $1$~keV/u solar wind energy, the He$^{2+}$-CO$_2$ reaction has a one-electron capture cross section of $5\times10^{-20}$~m$^2$ \citep{Greenwood2000,Bodewits2006}, whereas that of He$^{2+}$-CO is about $6\times10^{-20}$~m$^2$ \citep{Bodewits2006}. Because in H$_2$O, the $\sigma_{21}$ cross section is about $9\times10^{-20}$~m$^2$ (Paper~I), the difference in considering a H$_2$O-only atmosphere or a CO$_2$/CO one may lead to similar results, especially when deriving a total neutral outgassing rate from the in situ measurements of the He$^+$/He$^{2+}$ ratio (see Section~\ref{sec:inversion}). This conclusion holds for H$^+$ as well, as one-electron capture cross sections between protons and H$_2$O and CO$_2$ have the same magnitude, that is, about $20\times10^{-20}$~m$^2$ at $1$~keV/u impact energy \citep{Tawara1978,Greenwood2000}. 

\subsubsection{Variation with outgassing rate or heliocentric distance}\label{sec:outgassing}
The cometary water outgassing rate at a medium-activity comet such as 67P has been parameterized with respect to heliocentric distance by \cite{Hansen2016}, using the ROSINA neutral spectrometer on board \emph{Rosetta}. Depending on the inbound (pre-perihelion) and outbound (post-perihelion) legs, the total H$_2$O neutral outgassing rate $Q$ was
\begin{align}
        Q_\textnormal{in} & = (2.58\pm0.12)\times 10^{28}\ R_\textnormal{Sun}^{-5.10\pm0.05}\ \textnormal{s}^{-1} \label{eq:outgassingHansen} \\
        Q_\textnormal{out} & = (1.58\pm0.09)\times 10^{29}\ R_\textnormal{Sun}^{-7.15\pm0.08}\ \textnormal{s}^{-1},\nonumber
\end{align}

indicating an asymmetric outgassing rate with respect to perihelion. $R_\textnormal{Sun}$ denotes the heliocentric distance in AU. In order to obtain an estimate of the charge distribution of the solar wind during the \emph{Rosetta} mission, we chose to use the outgassing rate $Q_\textnormal{in}$ determined at inbound, where H$_2$O dominates the neutral coma. As mentioned above, close to the end of the mission, outside of $3$~AU, CO$_2$ became predominant \citep{Fougere2016,Lauter2018}.

We use in this section a Haser-like model \citep{Haser1957} including sinks, so that the density of water molecules $n_n$ at the comet is given by
\begin{align}
        n_n & = \frac{Q_\textnormal{in}}{4\pi \varv_0\ r^2} \exp\left(-\frac{f_\textnormal{d}\ (r-r_c)}{\varv_0}\right)\label{eq:HaserFull},
\end{align}
where $r$ is the cometocentric distance, $r_c$ is the comet's radius and $f_\textnormal{d}$ is the total photodestruction frequency (ionization plus dissociation) of H$_2$O as a result of the solar EUV flux. The effect of the exponential term becomes important at large cometocentric distances. $f_\textnormal{d}$ depends on the heliocentric distance \citep{Huebner2015}. In contrast to equation~(\ref{eq:HaserSimple}), which is valid for close orbiting around the comet, the exponential term is kept because of the large cometocentric distances considered here and the cumulative aspect of charge-changing reactions. $\varv_0$ is the radial speed of the neutral species, typically in the range $500-800$~m~s$^{-1}$ at comet 67P \citep{Hansen2016}. Speed $\varv_0$ is calculated using the empirically determined function of \cite{Hansen2016}:
\begin{align}
        \varv_0    &= \left(m_{R} R_\textnormal{Sun} + b_{R}\right)\ \left(1 + 0.171\ e^{-\frac{R_\textnormal{Sun}-1.24}{0.13}}\right),\label{eq:neutralSpeed}\\
    \textnormal{with}\quad m_{R} &= -55.5\quad\textnormal{and}\quad b_{R} =  771.0.
\end{align}
where $m_R$ and $b_R$ are fitting parameters, so that $\varv_0$ is expressed in m~s$^{-1}$.
The column density $\eta$ is integrated numerically.

In order to obtain an average effect, we chose here $\varv_0~=~600$\,m\,s$^{-1}$, $f_\textnormal{d} = 1.21\times10^{-5}\left(1 \textnormal{AU}/R_\textnormal{Sun}\right)^2$\,s$^{-1}$, corresponding to low solar activity conditions \citep[][including all photodissociation and photoionization channels]{Huebner2015}, and a constant solar wind bulk speed of $U_\textnormal{sw}=400$~\kms{}.

\begin{figure*}
  \includegraphics[width=\linewidth]{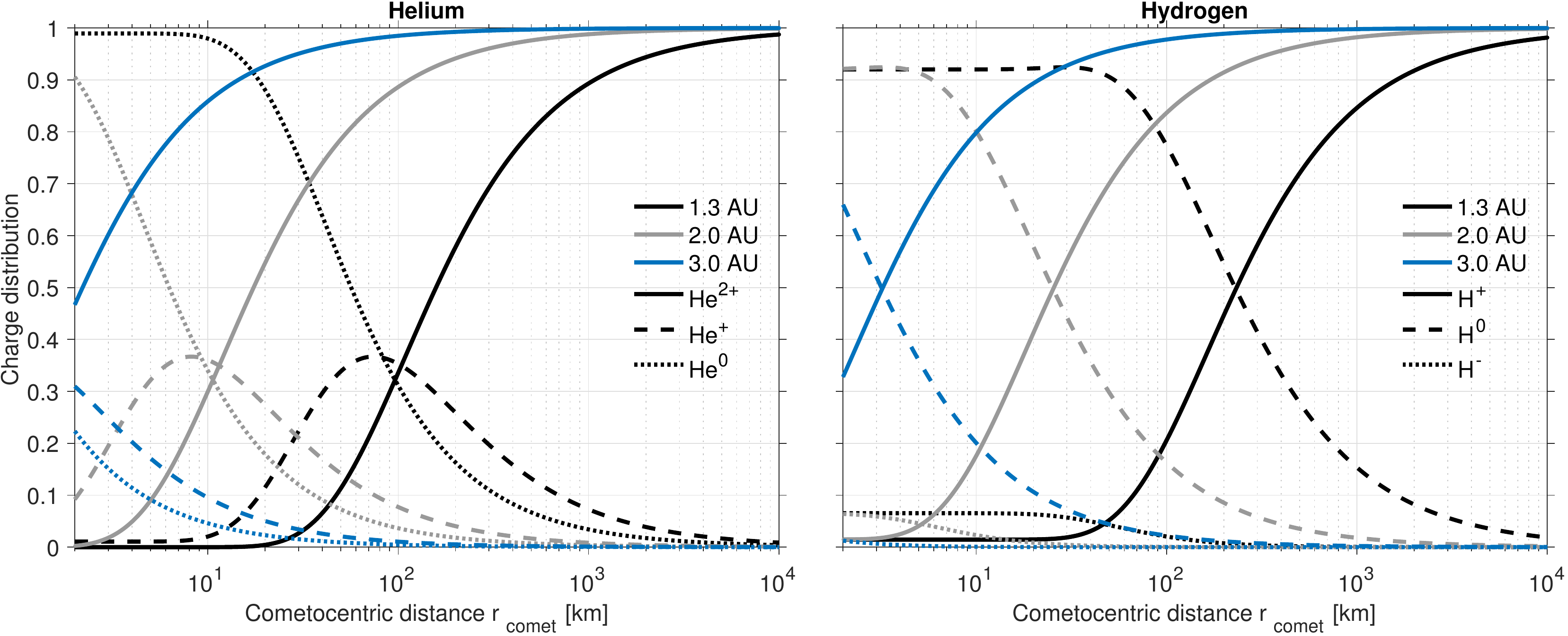}
     \caption{Normalized charge distributions of helium (left) and hydrogen (right) species in a H$_2$O 67P-like atmosphere for different outgassing rates (or heliocentric distances). The solar wind speed is assumed to be constant and equal to $400$~\kms{}.  (X$^{i+}$, X$^{(i-1)+}$, X$^{(i-2)+}$) components of projectile species X of initial positive charge $i$ are plotted as solid, dashed, and dotted lines, respectively.}
     \label{fig:ChargeFractionsHelio}
 \end{figure*}

Figure~\ref{fig:ChargeFractionsHelio} shows the beam fractionation for helium (left) and hydrogen (right) as a function of cometocentric distance, and for three heliocentric distances: $1.3$\,AU ($Q_0 = 6.8\times10^{27}$\,s$^{-1}$), $2$\,AU ($Q_0 = 7.5\times10^{26}$\,s$^{-1}$), and $3$\,AU ($Q_0 = 9.5\times10^{25}$\,s$^{-1}$). We note that the $1.3$\,AU case is only given here as a comparison to the other cases as it assumes no deflection of the incoming solar wind, no SWIC boundary formation, and thus is likely unrealistic \citep{Behar2016b,Behar2017}. Quasi-neutral hybrid plasma models are much better suited to realistically calculate these effects \citep{CSW2017}. That said, the validity of our model dependd on several parameters: the outgassing rate, solar EUV intensity, and solar wind parameters.  It also depends on the position of the spacecraft in a highly asymmetric plasma environment with respect to the Sun-comet line. All of these parameters may significantly fluctuate in a real-case \emph{Rosetta}-like scenario. This implies that the validity range of our model with respect to the cometocentric distance may extend or shrink depending on these parameters, and should thus be carefully evaluated in specific case studies. 

At $1.3$\,AU, as the solar wind approaches the comet nucleus and encounters a denser atmosphere, He$^{2+}$ become gradually converted into an equal mixture of He$^+$ ions and He$^0$ energetic neutral atoms, which become predominant below $100$\,km cometocentric distance. Correspondingly, all curves in Fig.~\ref{fig:ChargeFractionsHelio} (initially in black for $1.3$\,AU) are displaced toward smaller cometocentric distances with decreasing cometary outgassing rate and thus decreasing column density (gray and blue curves for $2$ and $3$\,AU). When \emph{Rosetta} was outside the SWIC region, that is, for $R_\textnormal{Sun}\gtrsim2$\,AU, its cometocentric distance was usually $10<r<100$\,km. For a constant solar wind speed of $400$\,\kms{}, this results in He$^{2+}$ being the most important helium species for most of the time during the solar wind ion measurements, with a proportion of about $\sfrac{1}{3}$ each for (He$^{2+}$, He$^+$, He$^0$) at the limit at $10$\,km cometocentric distance.

The hydrogen system presents a much simpler picture, with H$^-$ accounting for less than $7$\% of the total hydrogen beam at any cometocentric and heliocentric distances. At $1.3$\,AU, protons and H ENAs compose in equal parts the hydrogen beam at $200$\,km from the nucleus. Between $2$ and $3$\,AU, this balance occurs in the typical cometocentric distances explored by \emph{Rosetta}, that is, for $30$\,km distance and below. All scales considered, these conclusions are in qualitative agreement with those of \cite{Ekenback2008}, who used an MHD model to image hydrogen ENAs around the coma of comet 1P/Halley.

For reference, Appendix~\ref{appendix:collDepth} shows how the collision depth $\tau_i^\textnormal{cx}~=~\eta(r)~\sigma_i$ that is due to charge-changing processes in a H$_2$O atmosphere varies with cometocentric and heliocentric distances. It shows that the atmosphere is almost transparent to H and He ENAs, whereas H$^+$ and He$^{2+}$ ions will become much more efficiently charge-exchanged on their way to the inner coma.

As with the equilibrium charge states, charge distributions with a solar wind Maxwellian temperature of $T=3.6\times10^6$\,K and a solar wind bulk speed $U_\textnormal{sw}=400$\,\kms{} were also computed. Temperature effects are mostly seen for the hydrogen case, with an increase in loss cross sections from H$^0$, which are more efficiently converted back into H$^+$ and H$^-$ ions. Hence protons are not any more totally converted into their lower charge states when the solar wind becomes significantly heated. 

\subsubsection{Variation with solar wind speed}\label{sec:swspeed}
For our Sun, the solar wind speed varies typically between $300$ and $800$\,\kms{} and is not modified with increasing heliocentric distance \citep{Slavin1981}. The main variations are due to the regular (corotating interaction regions due to the Sun's rotation) and transient (coronal mass ejections) nature of the solar activity, and its subsequent dynamics in interplanetary space. In extreme cases, the solar wind speed, and thus the impact speed of the protons and $\alpha$ particles, may increase up to several thousand \kms{} in a matter of hours \citep{Ebert2009}. Adding solar wind temperature effects and heating at shock-like structures to these variations in bulk speed, large combined effects may arise in the charge distribution of solar wind particles.

\begin{figure*}
  \includegraphics[width=\linewidth]{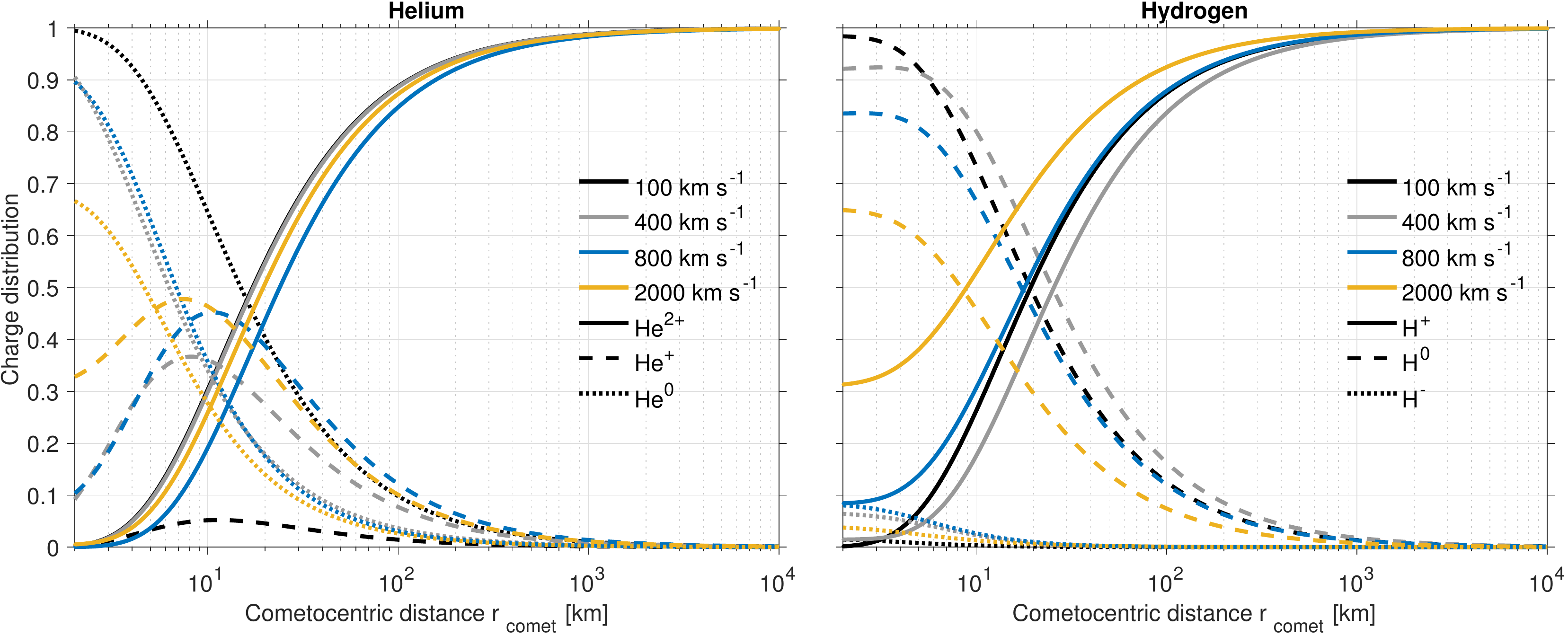}
     \caption{Normalized charge distributions of helium (left) and hydrogen (right) species in a H$_2$O 67P-like atmosphere for different solar wind speeds ($U_\textnormal{sw} = 100-2000$~\kms{}) and at a heliocentric distance of $2$~AU.}
     \label{fig:ChargeFractionsSWspeed}
 \end{figure*}

Figure~\ref{fig:ChargeFractionsSWspeed} shows the monochromatic charge distributions as a function of cometocentric distance for helium species (left) and for hydrogen species (right) for solar wind bulk speeds ranging between $100$ and $2000$~\kms{}. The calculations were made here for a distance of $2$~AU, hence at the limit when \emph{Rosetta} entered the SWIC; they are comparable to the gray curves in Fig.~\ref{fig:ChargeFractionsHelio}. A heliocentric distance of $2$~AU corresponds to a water outgassing rate of $Q_0 = 7.5\times10^{26}$~s$^{-1}$ and a neutral speed $\varv_0~\approx~600$~m~s$^{-1}$, chosen at inbound conditions \citep{Hansen2016}. 

At this level of cometary activity for 67P, no full-fledged bow shock structure is expected to have formed yet, although indications of a bow shock in the process of formation have been reported in \cite{Gunell2018} already at around $2.5$~AU and within $100$~km from the nucleus. These authors found that He$^{2+}$ ions move further downstream before being affected by the heating due to the presence of the shock-like structure. This implies that in this case, our model would be valid at lower cometocentric distances for helium particles than for hydrogen particles. Moreover, during these events, \emph{Rosetta} likely explored different locations in the comet-Sun plane containing the solar wind convection electric field because of the asymmetry of the bow-shock-like structure in this plane, hence modifying the validity range of the model depending on the off-$x$-axis position of the spacecraft. 
Therefore, our model is expected to be valid at a 67P-like comet down to typically a few tens of kilometers from the nucleus. Therefore, in this section, no thermal velocity distribution for the solar wind particles is assumed.

Owing to the velocity dependence of charge-changing reactions, a change in velocity in Fig.~\ref{fig:ChargeFractionsSWspeed} results for helium species in a complex behavior where the proportion of He$^{2+}$ ions (solid lines) first increases slightly from $100$ to $400$~\kms{}, decreases by about $10\%$ from $400$ to $800$~\kms{}, and finally increases again toward $2000$~\kms{} at any cometocentric distance to levels similar to those for $100$~\kms{}. In parallel, the proportion of He$^+$ (dashed lines) dramatically increases until about $800$~\kms{}, where it settles at a maximum around $45\%$ ($\sim10$~km cometocentric distance), a value that does not change much above this solar wind speed. This tendency can be more clearly seen in Fig.~\ref{fig:FractionsEffectProcesses} (left, black curves), where we calculate the charge distributions as a function of solar wind speed at $50$~km from the nucleus. Regarding He$^0$, it is interesting to note that the lower the solar wind speed, the larger the fraction of He$^0$ atoms. This is linked to the high double charge capture cross section of He$^{2+}$ at these energies as compared to the single charge capture, as discussed later in Section~\ref{sec:processes} \citep[see also][]{Bodewits2004}. At high impact speeds, this effect becomes reversed, and He$^+$ ions become relatively more abundant than He$^0$ atoms, and are the main loss channel of He$^{2+}$ ions.

For hydrogen species, the proportion of protons H$^+$ first diminishes ($10\%$ decline between $100$ and $400$~\kms{} on average) and then increases with solar wind speed in the $800-2000$~\kms{} range ($+30\%$ on average). The proportion of neutral atoms H$^0$ peaks below $400$~\kms{} solar wind speed; they may become dominant over H$^+$ at cometocentric distances below about $30$~km. These two effects are also shown in Fig.~\ref{fig:FractionsEffectProcesses} (right, black curves).
Similar to what we observed for the heliocentric distance study (Section~\ref{sec:swspeed}), H$^-$ ions make up only $10$\% or less of the solar wind, with a small increase seen below $10$~km cometocentric distance, where the atmosphere becomes increasingly denser; the maximum effect is reached when the solar wind bulk speeds are about $800$~\kms{}. 

\subsection{Role of double charge transfer and electron loss}\label{sec:processes}
We investigate now the effect of individual processes on the composition of the beam at a heliocentric distance of $2$~AU \citep[just outside of the SWIC, see][and previous section]{Behar2017}, and a cometocentric distance of $50$~km. The latter distance was chosen as a typical orbital distance of \emph{Rosetta} at $2$~AU. We used the recommended fitted monochromatic charge-changing cross sections of Paper~I, with solar wind speeds ranging from $100$~\kms{} to $5000$~\kms{}.

\begin{figure*}
  \includegraphics[width=\linewidth]{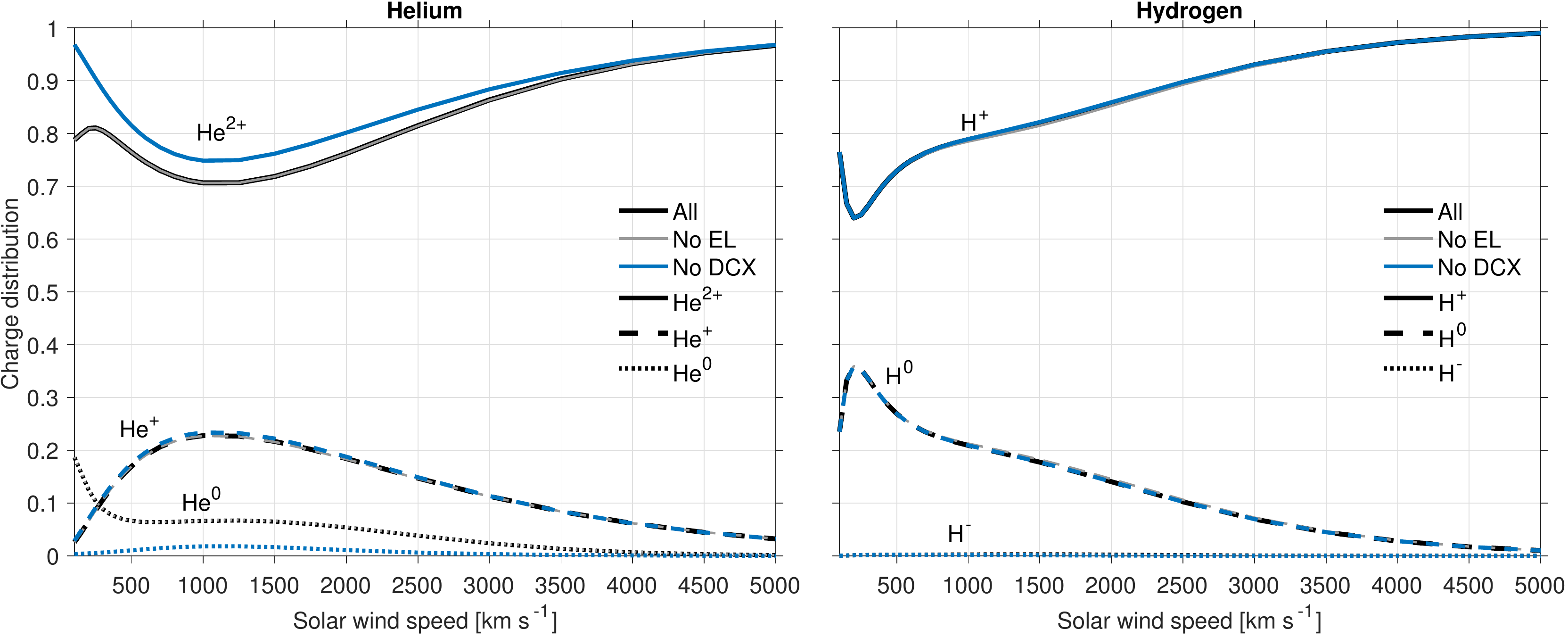}
     \caption{Normalized charge distributions of helium (left) and hydrogen (right) and effect of individual processes for a cometocentric distance of $50$~km and a heliocentric distance of $2$~AU. Double charge exchange (labeled "DCX") and electron loss (labeled "EL") cross sections are sequentially set to zero and compared to the full set (labeled 'All'). For hydrogen species, differences between all three runs are minimal. Solar wind speeds range from $100$~\kms{} to $5000$~\kms{}.}
 \label{fig:FractionsEffectProcesses}
 \end{figure*}

Figure~\ref{fig:FractionsEffectProcesses} shows the charge distribution of helium (left) and hydrogen (right) species as a function of solar wind speed at $50$~km cometocentric distance.
From Fig.~9 of Paper~I, double charge exchange (DCX) He$^{2+}\rightarrow$He$^0$ is expected to be the main loss of $\alpha$ particles at solar wind speeds below $300$~\kms{}, leading to the creation of He$^0$ atoms, whereas single charge exchange He$^{2+}\rightarrow$He$^+$ starts to play a more important role at higher solar wind speeds. When we set the DCX cross sections to zero ($\sigma_{20} = 0$ and $\sigma_{02} = 0$) in our simulations (Fig.~\ref{fig:FractionsEffectProcesses}, blue curves), the solar wind contains less than $2$\% He$^0$ atoms at any impact speed, whereas their proportion climbs up to almost $20$\% at $100$~\kms{} when DCX is taken into account. As expected from the shapes of the cross sections and the relative abundance of He$^{2+}$ and He$^0$, the most important effect is for the $2\rightarrow0$ charge process.
We also study how electron loss (EL) processes ($\sigma_{01}$, $\sigma_{02}$, $\sigma_{12}$) impact the charge distributions with respect to solar wind speed. This is shown as gray curves in Fig.~\ref{fig:FractionsEffectProcesses}. No drastic change is seen when the EL processes are turned on or off in our simulations (black and gray curves are almost superimposed in this figure).

For hydrogen species, neither EL nor DCX processes seem to play any significant role in the composition of the beam at $2$~AU, implying that the main processes populating all three species at the comet are single-electron capture. This analysis is further vindicated by the behavior of the hydrogen system with respect to heliocentric and cometocentric distances (see Sections~\ref{sec:outgassing} and \ref{sec:swspeed}). That said, EL processes may start playing a role at solar wind speeds above $800$~\kms{} and for cometocentric distances below about $10$~km, where the neutral column density becomes comparatively much higher.

Maxwellian-averaged cross sections can also be used here; because DCX usually peaks at low impact velocities, He$^{2+}$ ions will be less efficiently converted into He atoms with increasing solar wind temperature. Differences in the charge composition of the solar wind, especially below $300$\kms{} , will start to appear (figure not shown) for temperatures $T\gtrsim5$~MK for helium particles (relative increase in He$^{2+}$ and He$^+$ over He$^0$), and for $T\gtrsim8$~MK for hydrogen particles (relative increase in H$^+$ over H$^0$). 

Because EL processes are expected to play a minor role at \emph{Rosetta}'s position around comet 67P, flux charge distributions and arguably simpler expressions for the reduced EL-free system can be derived. These equations are presented in Appendix~\ref{appendix:reducedSystem} for clarity.

\subsection{Simulated charge distribution during the \emph{Rosetta} mission}\label{sec:syntheticCX}

To finalize our theoretical study of charge-changing processes at a 67P-like comet, we now first turn to evaluating the normalized composition of the solar wind helium and hydrogen charge distributions in the vicinity of 67P throughout the \emph{Rosetta} mission (2014-2016).
Using the analytical model inversions presented in Section~\ref{sec:inversion}, we then show how the outgassing rate and solar wind upstream fluxes can be reconstructed from the in situ knowledge of the He$^+$-to-He$^{2+}$ ratio and proton flux, and we apply this technique to the complex trajectory of \emph{Rosetta} around comet 67P. Validations of these inversions are presented in Sections~\ref{sec:syntheticOutgassing} and \ref{sec:syntheticSW} and follow the following scheme: $(i)$ generate virtual measurements from basic upstream solar wind and cometary outgassing rate parameters, $(ii)$ use forward analytical model to produce the expected solar wind charge distributions locally at the geometric position of \emph{Rosetta}, and $(iii)$ perform inversions from the locally generated fluxes to retrieve the upstream solar wind conditions or the outgassing rate.

\subsubsection{Solar wind composition}

This paragraph aims at simulating what an electron-ion-ENA spectrometer would observe at the location of \emph{Rosetta} around comet 67P.
Because of the large dataset that we attempt to simulate, we derived here the column density $\eta$ following the simple 2D integration of \cite{Beth2016}, and set the exponential term in equation~(\ref{eq:HaserFull}) to one. At the position of \emph{Rosetta}, the difference between including or excluding the exponential loss term that is due to photodestruction is negligible, as discussed in Section~\ref{sec:inversion}. The column density is given by equation~(\ref{eq:columnDensity}) with the neutral outgassing rate and speed $\varv_0$ parameterized by equations~(\ref{eq:outgassingHansen}) and (\ref{eq:neutralSpeed}) \citep[see][]{Hansen2016}. 
The geometry of \emph{Rosetta} in different coordinate systems, including CSEq, is accessible via the European Space Agency Planetary Science Archive (PSA).
For simplicity, an average solar wind speed of $400$~\kms{} \citep{Slavin1981} was chosen to calculate the total charge-changing cross sections during the mission. Solar wind propagation from point measurements at Mars (with the Mars Express, MEX, spacecraft) and at Earth (with the ACE satellite) would provide a more physically accurate upstream solar wind, although at the expense of simplicity in our theoretical interpretation. For the comparison with \emph{Rosetta} observations, we refer to Paper~III. 

\begin{figure*}
  \includegraphics[width=\linewidth]{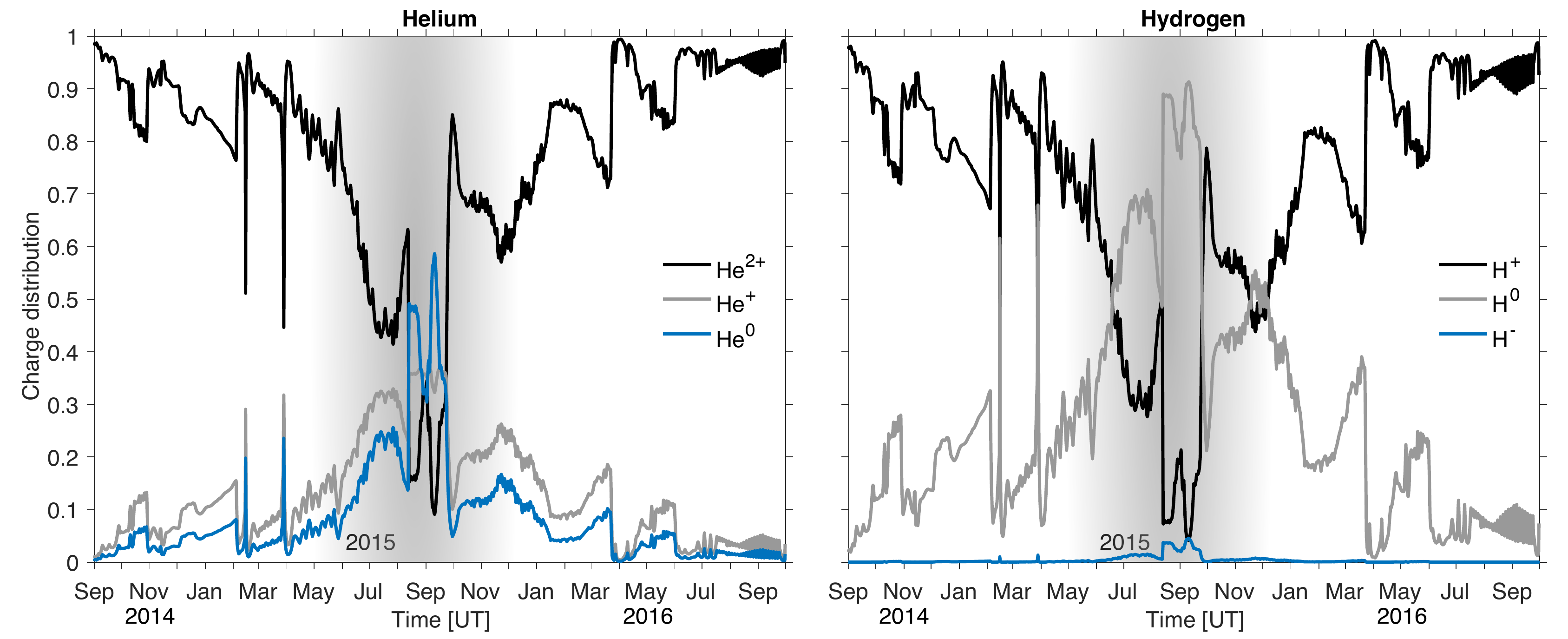}
     \caption{Expected normalized charge distributions of helium (left) and hydrogen (right) during the \emph{Rosetta} mission (2014-2016) at the location of the spacecraft. The SWIC encountered at comet 67P is marked as a gray region, with a smooth gradient to indicate its dynamic nature.}
 \label{fig:FractionsRosettaMission}
 \end{figure*}

Figure~\ref{fig:FractionsRosettaMission} shows the simulated normalized charge distributions at the position of \emph{Rosetta} for comet 67P between 2014-2016 for helium (left) and hydrogen (right) species. The SWIC, where the solar wind was mostly prevented from entering the inner coma, is shown as a gray-graded region and spans almost $\text{eight}$ months between late April and early December 2015 \citep{Behar2017}. It corresponds to times where the column density traversed by the solar wind beams becomes comparatively high. 

In this ion cavity, both He$^{2+}$ and H$^+$ beams are strongly depleted in favor of lower charge states, which coincides with the lack of in situ observations during this period \citep{Behar2017}. Whether this cavity has a well-defined surface, or how dynamical it is (with regard to the spacecraft position), are questions that are unanswered as of now because they challenge the ion sensors at the limit of their capacity (field-of-view limitations and sensitivity).
Our analytical model does not take into account the complex trajectories of solar wind particles in the inner coma \citep[as pointed out in][]{Behar2018aa_model,Saillenfest2018}, which is likely to increase the efficiency of the CX because of the curvilinear path of projectiles, which also depends on their charge and mass. 
Consequently, the correct origin of the SWIC may be better investigated by a self-consistent modeling that includes the physico-chemistry of the coma, such as a quasi-neutral hybrid plasma model \citep{Koenders2015,CSW2017,Lindkvist2018}. In our results, the analytical calculations should in this region only be seen as an indication of the charge distribution of the solar wind for rectilinear trajectories of the incoming solar wind.

For the helium system, He$^{2+}$ constitutes the bulk of the charged states, reaching percentages of at least $70$\% outside of the SWIC. He$^+$ ions and He$^0$ atoms have a similar behavior and are each about $15$\% of the total helium solar wind. Because of the changing geometry and outgassing rate, the compositional fractions are asymmetric with respect to perihelion. Because no ENA detector was on board \emph{Rosetta}, the full charge distribution of the solar wind cannot be determined; a new mission to another comet could thus usefully include such an instrument \citep{Ekenback2008}. In two instances before perihelion, in February 2015 and at the end of March 2015 ($R_\textnormal{Sun}\sim$2.5~AU), the fractions of He$^0$ and He$^+$ increased dramatically after the spacecraft orbited within $10$~km from the nucleus. As shown in Fig.~\ref{fig:ChargeFractionsHelio}, the proportion of these two charge states increases dramatically in and around this cometocentric distance and closely matches that of He$^{2+}$ ions.

For the hydrogen system, in a way similar to the helium system, the solar wind contains mostly protons, with an average percentage of $70$\% outside of the SWIC. In the two instances described above, H$^0$ atoms also become more abundant than H$^+$. In agreement with the previous sections, hydrogen negative ions H$^-$ only seem to be of note around perihelion, where it reaches about $3-4\%$ of the total (outside of the SWIC, the abundance levels are closer to $0.1\%$). 

Negative hydrogen ions were first discovered by \cite{Burch2015} early in the mission and up to January 2015, using the RPC-IES electron instrument on board \emph{Rosetta}. \cite{Burch2015} ascribed the observed H$^-$ to the two-step charge-transfer process $\textnormal{H}^+\rightarrow\textnormal{H}^0\rightarrow\textnormal{H}^-$ from solar wind protons around $1$~keV energy ($\sim437$~\kms{}). At this bulk speed, our values of $\sigma_{10}$ ($1.7\times10^{-19}$~m$^{2}$) and $\sigma_{0-1}$ ($6.4\times10^{-21}$~m$^{2}$) are similar within a factor $2$ to those used by \cite{Burch2015}, whereas our value of $\sigma_{1-1}$ ($4.6\times10^{-23}$~m$^{2}$) is a factor $3.2$ lower (these authors used values for Ar and O$_2$ for this reaction). Consequently, their main conclusions remain unchanged: the two-step process is in our new calculations about $23$ times more efficient than DCX reactions to produce H$^-$ anions. When making the numerical application and correcting their two-step process to $1\times10^{-3}~F^\textnormal{sw}$, and of double capture to $3\times10^{-4}~F^\textnormal{sw}$, \cite{Burch2015} should have found a ratio of about $3.4$.

The simulated H$^-$ component in our simulation is very faint and therefore points to the presence of favorable neutral-plasma conditions (increased outgassing, small cometocentric distance, increased solar wind flux, or combinations thereof) in order to be detectable. This conclusion is contained in the account of \cite{Burch2015}.

\subsubsection{Outgassing rate retrieval}\label{sec:syntheticOutgassing}

This section aims at validating our inversion procedure for the outgassing rate, using the He$^+$/He$^{2+}$ ratio as a proxy of the neutral outgassing at the comet. We follow three steps: $(i)$ calculation of the He$^+$/He$^{2+}$ ratio at \emph{Rosetta} during the mission, using the forward analytical model with the neutral atmosphere of \cite{Hansen2016} as inputs (as in Fig.~\ref{fig:FractionsRosettaMission}), $(ii)$ computation of the geometric factor $\epsilon(r,\chi)$ entering in the expression of the column density (see equation~(\ref{eq:columnDensity})), which depends on \emph{Rosetta}'s position around comet 67P during the mission, $(iii)$ final reconstruction of the outgassing rate from the local He$^+$/He$^{2+}$ ratio, using equation~(\ref{eq:outgassingRetrieved}).

\begin{figure*}
  \includegraphics[width=\linewidth]{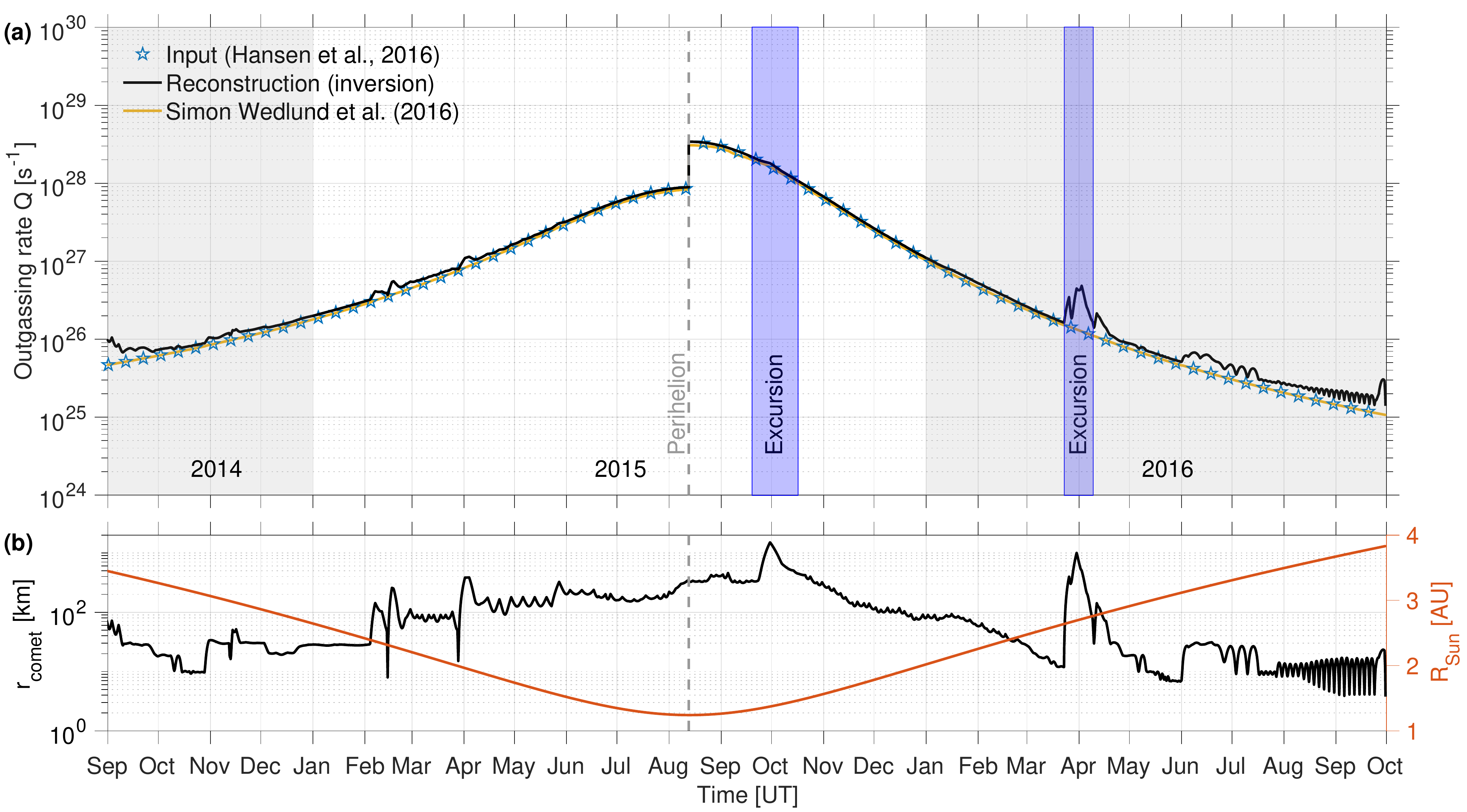}
     \caption{(a) Outgassing rate $Q_0$ reconstructed from the inversion of the analytical model (black line), compared to the original input outgassing rate \citep{Hansen2016} (blue stars) during the \emph{Rosetta} mission (2014-2016). The orange line is the result of the simplified approach of equation~(\ref{eq:outgassingCSW2016}) where only $\textnormal{He}^{2+}\rightarrow\textnormal{He}^+$ reactions were considered \citep[][]{CSW2016}; this method works well at the constant solar wind speed of $400$~\kms{} chosen here. (b) Geometry parameters of the spacecraft \emph{Rosetta} at comet 67P during its two-year mission: black, cometocentric distance; red, heliocentric distance. Two large cometocentric distance excursions are indicated in blue: the dayside excursion (September-October 2015), and the cometary tail excursion (March-April 2016). Gray-shaded regions mark years.}
 \label{fig:OutgassingRateSynthetic}
 \end{figure*}

Figure~\ref{fig:OutgassingRateSynthetic} presents the results of this approach and compares our reconstructed outgassing rate (black) with the production rate fit of \cite{Hansen2016}, which was used in the first place to generate the charge distributions in Fig.~\ref{fig:FractionsRosettaMission}. Very good agreement within $15\%$ on average is found throughout the mission, except for occasional events, such as the cometary tail excursion around April 2016 or at the end of the mission. This stems from the approximation made in the inversion procedure detailed in Section~\ref{sec:inversion}, with the condition
$F_i^\infty/F_i \ll 1,\ \textnormal{for}\ i=1,2$. This condition is fulfilled for He$^{2+}$ but not for He$^+$ during the tail excursion (because of the large cometocentric distance and comparatively low column density), and during the early and later parts of the mission (very low column density for a comparatively small cometocentric distance). 
The sweet spot of the retrieval method with the fulfilled condition for He$^+$ is consequently achieved in the region where the He$^+$ charge fraction is peaking with respect to the cometocentric distance (see this region in  Fig.~\ref{fig:ChargeFractionsHelio}, left, dashed lines). Outside of these regions, and if the solar wind speed is about $400$~\kms{} or above, a much simpler approach, as detailed in \cite{CSW2016} and epitomized by equation~(\ref{eq:outgassingCSW2016}), may prove better suited. This is demonstrated by the yellow line in Fig.~\ref{fig:OutgassingRateSynthetic}, which at this constant solar wind speed agrees with the input outgassing rate of \cite{Hansen2016} to within $5\%$. However, during the \emph{Rosetta} mission, this solar wind speed value is only encountered episodically, as can be seen in the solar wind velocities measured by the RPC-ICA ion spectrometer \citep{Behar2017} and the more complex approach developed in the present study, with $\text{six}$ charge-changing cross sections, is warranted.

 \subsubsection{Solar wind upstream retrieval}\label{sec:syntheticSW}
The second inversion introduced in Section~\ref{sec:inversion} enables reconstructing the upstream solar wind flux or density from local measurements made deep into the coma. To test our inversion, we first created synthetic upstream solar wind conditions, which we propagated with the forward analytical model at the position of \emph{Rosetta}.

According to the parameterization of \cite{Slavin1981} with respect to heliocentric distance, the undisturbed proton density is $n_p = 7\times10^6\,R_\textnormal{Sun}^{-2}$~m$^{-3}$, with $R_\textnormal{Sun}$ expressed in AU. For a constant $400$~\kms{} solar wind bulk speed, this is equivalent to an upstream solar wind proton flux $F_p = 2.8\times10^{12}\,R_\textnormal{Sun}^{-2}$\,m$^{-2}$\,s$^{-1}$, which is commensurable to the flux levels measured by the RPC-ICA instrument on board \emph{Rosetta} \citep{Nilsson2017b,Nilsson2017a}. On average, the solar wind is composed of about $4\%$ He$^{2+}$ ions \citep[e.g.,][]{CSW2017}. We first applied the analytical model to the inputs above and calculated the resulting local proton and helium ion fluxes at the position of \emph{Rosetta} during the mission; this is equivalent to multiplying the normalized charge distributions in Fig.~\ref{fig:FractionsRosettaMission} by the upstream solar flux $F_p$ for protons, and by $F_\alpha=\sfrac{1}{24}\,F_p$ for $\alpha$ particles. Using equation~(\ref{eq:inversionSW}), we then reconstructed the upstream solar wind flux from the synthetic fluxes. 

\begin{figure*}
  \includegraphics[width=\linewidth]{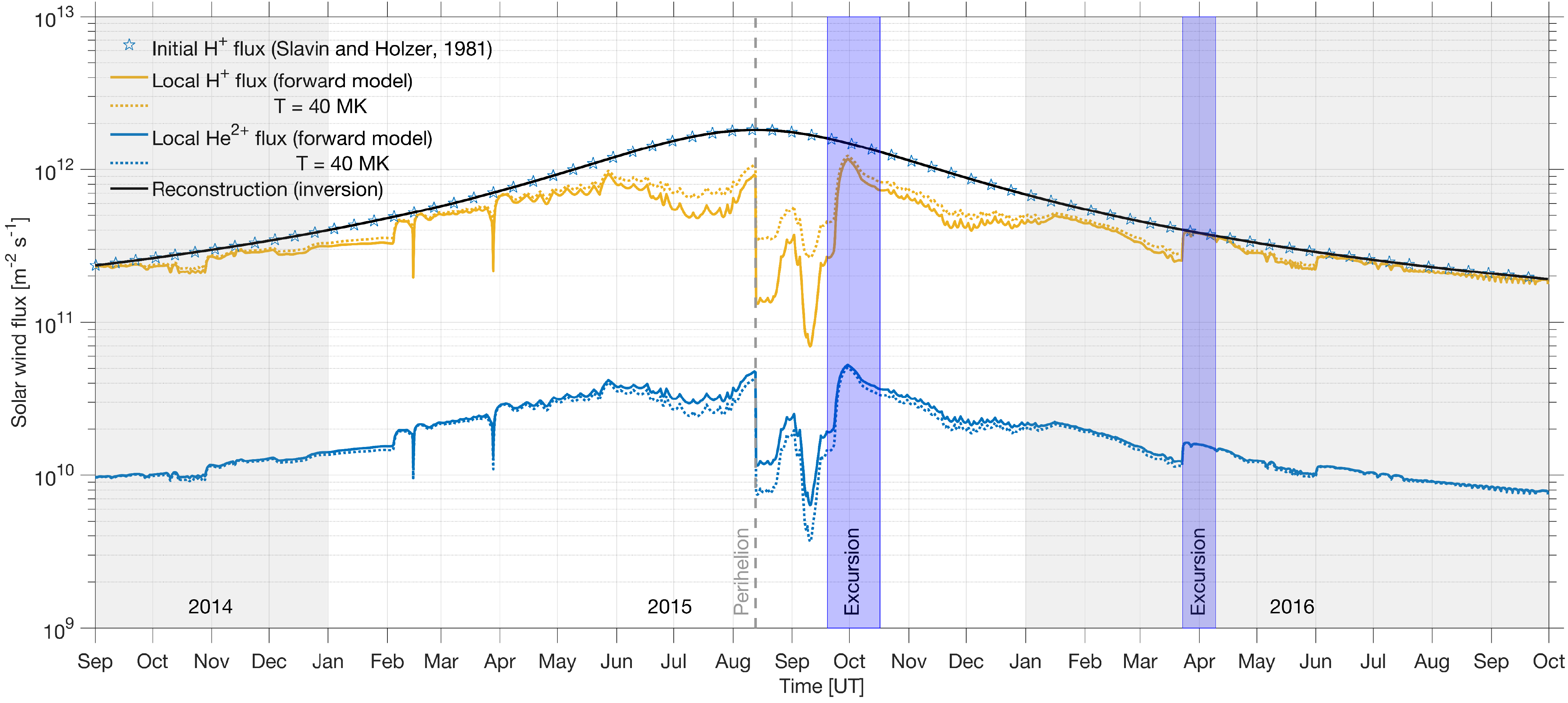}
     \caption{Solar wind flux reconstructed from the inversion of the analytical model compared to the original input solar wind upstream flux \citep{Slavin1981} during the \emph{Rosetta} mission (2014-2016). Proton and $\alpha$ particle fluxes ($4\%$ of the total solar wind ions) are considered. Dotted lines are results of the forward analytical model with a solar wind Maxwellian temperature $T=40\times10^6$~K. Otherwise, the caption is the same as for Fig.~\ref{fig:OutgassingRateSynthetic}.}
 \label{fig:SolarWindSynthetic}
 \end{figure*}

The results are presented in Fig.~\ref{fig:SolarWindSynthetic}, where the solar wind input flux and the reconstructed upstream flux match perfectly. In conformity with Fig.~\ref{fig:FractionsRosettaMission}, the solar wind fluxes are expected to decrease by almost one order of magnitude around perihelion at the position of \emph{Rosetta} as a result of CX processes.

For comparison purposes, we also calculated the effect of very high Maxwellian temperatures for the solar wind, with $T = 40\times10^6$\,K reminiscent of a strong heating at a full-fledged bow shock structure in the upstream solar wind. These temperatures correspond to thermal velocities of $1000$\,\kms{} for protons and $500$\,\kms{} for $\alpha$ particles. This is shown in Fig.~\ref{fig:SolarWindSynthetic} as dotted lines. Increasing the temperatures leads to a similar trend, albeit reinforced, to the trend that we previously described in Section~\ref{sec:EqChargeDistribution}: proton fluxes are reinforced (factor $\times3.8$ at perihelion), especially in the so-called SWIC region, whereas He$^{2+}$ fluxes undergo a decrease by a factor of about $1.8$ at perihelion.
As pointed out in the sections above, the use of measured solar wind fluxes, propagated from Mars and Earth observations to \emph{Rosetta}'s position \citep[as in][]{Behar2018thesis}, and how they connect with the local flux measurements made with RPC-ICA, will be discussed in our next study (Paper~III). 

\section{Conclusions}
We have developed a 1D analytical model of charge-changing reactions at comets based on the fluid continuity equation and within the assumptions of stationarity and of particle motion along solar wind streamlines at the same bulk speed.
A sensitivity study on several cometary parameters was then conducted for helium and hydrogen three-component systems. The results are listed below.
\begin{itemize}
\item Double charge transfer is important for helium, especially at solar wind velocities below about $500$~\kms{}.
\item Electron loss (stripping) plays only a minor role in the composition of the solar wind at any solar wind impact speed and at the typical cometocentric and heliocentric distances encountered by the \emph{Rosetta} spacecraft. For high solar wind speeds ($>800$\,\kms{}) and much higher column densities, stripping effects may start to appear, especially for hydrogen projectiles. 
\item Solar wind temperature effects start to play a role at temperatures $T>3\times10^6$~K, in accordance with \cite{CSW2018a}. At comet 67P at the position of \emph{Rosetta}, this results in an increase in proton fluxes by a factor $3-4$ around perihelion, whereas $\alpha$ particles are further depleted compared to a monochromatic (monoenergetic) solar wind.    
\end{itemize}

We have also shown that with this analytical model,  the charge-state distribution of helium and hydrogen species in cometary atmospheres can be predicted, with the use of a total of $12$ charge-changing reactions in a water atmosphere \citep[see][for recommended cross sections]{CSW2018a}. We predict at a 67P-like comet the formation of a region below $2$~AU where the incoming solar wind ions are efficiently lost to lower charge states and ENAs through CX reactions alone. In combination with kinetic plasma effects and the formation of shock-like structure upstream of the nucleus \citep{Gunell2018}, CX may thus play an additional role in the creation of the solar wind ion cavity characterized with \emph{Rosetta} by \cite{Behar2017}. From the knowledge of in situ ion composition applied to He$^+$ and He$^{2+}$, we also demonstrated that it is possible to retrieve the outgassing rate of neutrals and solar wind upstream conditions purely from geometrical considerations and from local measurements made deep into the coma, assuming a spherically symmetric $1/r^2$ expansion for the neutral atmosphere.

This article is the second part of a triptych on charge-transfer efficiency around comets. The first part gives recommendations on low-energy charge-changing and ionization cross sections of helium and hydrogen projectiles in a water gas. The third part, presented in \cite{CSW2018c}, aims at applying this analytical model and its inversions to the Rosetta Plasma Consortium (RPC) datasets, and in doing so, at quantifying charge-transfer reactions and comparing them to other processes during the \emph{Rosetta} mission to comet 67P.

\appendix

\section{Hydrogen system, forward model} \label{appendix:HydrogenTheory}
We present here the explicit analytical model for the system of (H$^+$,~H$^0$,~H$^{-}$) and its $\text{six}$ charge-changing reactions with a neutral atmosphere. The solution is identical to that presented in Section~\ref{sec:Helium} for the helium system and is given here for completeness in the manner of \cite{Allison1958}. The relevant cross sections are in this case
\begin{align*}
        \sigma_{10}:&\ \textnormal{H}^{+} \longrightarrow \textnormal{H}^0 &\ \textnormal{single capture} \\
    \sigma_{1-1}:&\ \textnormal{H}^{+} \longrightarrow \textnormal{H}^{-} &\ \textnormal{double capture}\\    
    \sigma_{01}:&\ \textnormal{H}^{0} \longrightarrow \textnormal{H}^{+} &\ \textnormal{single stripping} \\
    \sigma_{0-1}:&\ \textnormal{H}^{0} \longrightarrow \textnormal{H}^{-} &\ \textnormal{single capture} \\
    \sigma_{-11}:&\ \textnormal{H}^{-} \longrightarrow \textnormal{H}^{+} &\ \textnormal{double stripping}\\
    \sigma_{-10}:&\ \textnormal{H}^{-} \longrightarrow \textnormal{H}^{0} &\ \textnormal{single stripping}
\end{align*}
 
We may correspondingly pose
\begin{equation*}
        \begin{split}
                \sigma_1 &= \sigma_{10} + \sigma_{1-1} & \textnormal{for H}^{+}\\ 
        \sigma_0  &= \sigma_{01} + \sigma_{0-1} & \textnormal{for H}^{0}\\
        \sigma_{-1} &= \sigma_{-11} + \sigma_{-10} & \textnormal{for H}^{-} \\
        \sum{\sigma_{ij}} &= \sigma_1 + \sigma_0 + \sigma_{-1},
        \end{split}
\end{equation*}
with $\sum{\sigma_{ij}}$ the sum of all six cross sections. 

For an upstream solar wind flux $F^\textnormal{sw}$, and with $F_1$, $F_0$ and $F_{-1}$ representing H$^{+}$, H$^{0}$ and H$^{-1}$ fluxes, the matrix system~(\ref{eq:matrixsysGeneral}) of equations for the reduced (H$^+$,~H$^0$) system with $N=2$ is
\begin{align}
  \frac{\drv}{\drv\eta}  \begin{bmatrix} F_1 \\ F_0 \end{bmatrix}  &= 
  \begin{bmatrix} a_{11} & a_{10}\\ a_{01} & a_{00} \end{bmatrix} 
  \begin{bmatrix} F_1 \\ F_0 \end{bmatrix} 
  +F^\textnormal{sw}\begin{bmatrix} \sigma_{-11} \\ \sigma_{-10} \end{bmatrix},\label{eq:hydrogen_simple}\\
  \textnormal{with}&\nonumber\\
  \begin{split} 
                a_{11} &= -(\sigma_1 + \sigma_{-11}), \qquad a_{10} = \sigma_{01} - \sigma_{-11}, \\
        a_{01} &= \sigma_{10} - \sigma_{-10}, \qquad  a_{00} = -(\sigma_0 + \sigma_{-10}).
        \end{split}\nonumber
\end{align}

Solution~(\ref{eq:HeliumSolution}) for the helium system can be made to apply to the hydrogen system by subtracting every finite index of flux and cross section by $1$, so that
\begin{align}
     \vec{F} &= F^\textnormal{sw} \left(\vec{F}^{\infty} + \frac{1}{2q}\left( \vec{P}\ e^{q \eta} - \vec{N}\ e^{-q \eta} \right)\ e^{-\frac{1}{2} \sum{\sigma_{ij}}\,\eta}\right)\label{eq:HydrogenSolution}\\
    \textnormal{with} \nonumber\\
    \begin{split}
         \vec{F}^\infty &= \begin{bmatrix} F_1^\infty \\ F_0^\infty \\ F_{-1}^\infty \end{bmatrix} \\
         &= \frac{1}{D}\begin{bmatrix}  -a_{00}\sigma_{-11}+a_{10}\sigma_{-10} \\\left(a_{01}\sigma_{-11} - a_{11}\sigma_{-10}\right)\\ \resizebox{0.36\textwidth}{!}{$\sigma_{-11}(a_{00}-a_{01}) + a_{11}(a_{00}+\sigma_{-10}) - a_{10}(a_{01} + \sigma_{-10})$}\end{bmatrix},\\
        \vec{P} &= \begin{bmatrix} P_1 \\ P_0 \\ P_{-1} \end{bmatrix} = \begin{bmatrix} (t + q)\left(1 - F_1^\infty\right) - a_{10} F_0^\infty\\ a_{01}\left(1 - F_1^\infty\right) + (t - q) F_0^\infty \\ -(t+q+a_{01})\left(1 - F_1^\infty\right) - (t-q-a_{10}) F_0^\infty  \end{bmatrix},\\
        \vec{N} &= \begin{bmatrix} N_1 \\N_0 \\ N_{-1} \end{bmatrix} = \begin{bmatrix} (t - q)\left(1 - F_1^\infty\right) - a_{10} F_0^\infty\\ a_{01}\left(1 - F_1^\infty\right) + (t + q) F_0^\infty\\ -(t-q+a_{01})\left(1 - F_1^\infty\right) - (t+q-a_{10}) F_0^\infty  \end{bmatrix},\\
        \textnormal{recalling} \\
        t &= \frac{1}{2} (a_{11}-a_{00}), \quad q =\frac{1}{2}\sqrt{(a_{11}-a_{00})^2 + 4a_{01}a_{10}},\\
        \sum{\sigma_{ij}} &= -(a_{00} + a_{11}),\quad\textnormal{and}\quad  D = a_{00}a_{11}-a_{01}a_{10}.
        \end{split}\nonumber
\end{align}

\section{Electron loss-free helium system and simplified formula for outgassing rate}\label{appendix:reducedSystem}

\begin{figure*}
  \includegraphics[width=\linewidth]{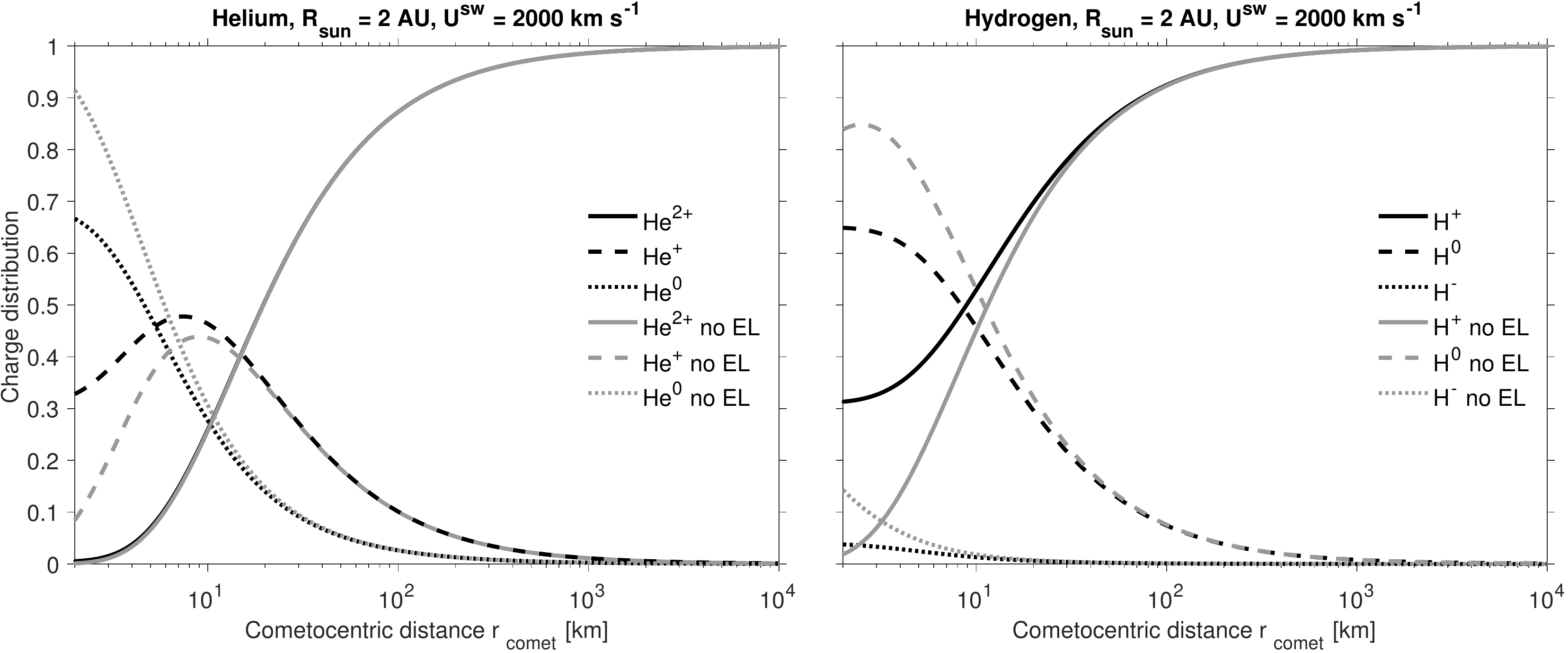}
     \caption{Normalized charge distributions of helium (left) and hydrogen (right) with respect to cometocentric distance for a solar wind speed of $2000$~\kms{} and a heliocentric distance of $2$~AU. The distributions are analytically calculated with and without electron loss reactions (EL).}
     \label{fig:CX_noEL}
 \end{figure*}

As shown in Section~\ref{sec:discussion}, electron loss reactions do not play a significant role at typical solar wind speeds and for the heliocentric distances encountered during the orbiting phase of \emph{Rosetta}.
Ignoring the three electron loss reactions $\sigma_{12}$, $\sigma_{01}$ , and $\sigma_{02}$, that is, with only CX reactions considered, the flux continuity equation for the (He$^{2+}$, He$^+$, He$^0$) helium system with fluxes ($F_2$, $F_1$, $F_0$) reduces to
\begin{equation}
        \begin{cases}
                \begin{split}
                        \frac{\drv F_2}{\drv\eta}   &= -\left(\sigma_{21}+\sigma_{20}\right)\ F_2\\
                \frac{\drv F_1}{\drv\eta}    &= \sigma_{21}\ F_2 - \sigma_{10}\ F_1,\\
                \frac{\drv F_0}{\drv\eta} &= \sigma_{20}\ F_2 + \sigma_{10}\ F_1
                \end{split}\label{eq:helium_ce_reduced}
     \end{cases}
\end{equation}
where, by definition, $F_0 = F^\mathrm{sw} - \left(F_2 + F_1\right)$. Solving this system of single differential equations, we find the expression of fluxes depending on column density $\eta$:
\begin{equation}
        \begin{cases}
                \begin{split}
                        F_2   &= F^\mathrm{sw}\ e^{-\left(\sigma_{21}+\sigma_{20}\right)\,\eta}\\[0.25em]
                F_1   &= \frac{\sigma_{21}}{\sigma_{10}-\left(\sigma_{21}+\sigma_{20}\right)}\ \left(F_2 - F^\mathrm{sw}\,e^{-\sigma_{10}\,\eta}\right),\\[0.25em]
                F_0   &= F^\mathrm{sw} - \left(F_2 + F_1\right)
                \end{split}\label{eq:helium_reduced}
     \end{cases}
\end{equation}
which is considerably simpler than the full $\text{six}$-reaction solution~(\ref{eq:HeliumSolution}). This expression yields results that are almost identical to the full $\text{six}$-reaction model in the conditions probed by \emph{Rosetta}. Differences between the two approaches are negligible at solar wind speeds of $400$\,\kms{} and below, but may become noticeable for higher values, when the maxima of stripping cross sections are approached. An illustration of the difference expected at comet 67P between the $\text{six}$-reaction model and the present electron loss-free solution is shown in Fig.~\ref{fig:CX_noEL} at $2$\,AU for a solar wind speed of $2000$\,\kms{}. Such high speeds can be encountered in extreme solar transient events such as coronal mass ejections \citep{Meyer2012}. In this case, electron loss reactions start to play a role below $10$~km cometocentric distance for helium and below about $30$~km for hydrogen. 

We may extract the column density $\eta$, which depends on cometocentric distance $r$ and solar zenith angle $\chi$, by calculating the flux ratio $\mathcal{R}=F_1/F_2$ as in Section~\ref{sec:inversion}:
\begin{align}
     \eta(r,\chi) & = \frac{\ln\left(1+ \frac{\sigma_{20}+\sigma_{21}-\sigma_{10}}{\sigma_{21}}\,\mathcal{R}\right)}{\sigma_{21}+\sigma_{20}-\sigma_{10}}. 
\end{align}

Taking the definition of the approximate cometary neutral column density from equation~(\ref{eq:columnDensity}), that is, $\eta(r,\chi)~=~\frac{Q_0}{4\pi \varv_0\, r} \frac{\chi}{\sin\chi}~=~\frac{Q_0}{\varv_0}\, \epsilon(r,\chi)$, the cometary neutral outgassing rate is under these assumptions
\begin{align}
    Q_0 & = \frac{\varv_0}{\epsilon(r,\chi)}\  \frac{\ln\left(1+ \frac{\sigma_{20}+\sigma_{21}-\sigma_{10}}{\sigma_{21}}\,\mathcal{R}\right)}{\sigma_{21}+\sigma_{20}-\sigma_{10}}.
\end{align}

This expression reduces further to equation~(\ref{eq:outgassingCSW2016}) when we pose $\sigma_{20} = \sigma_{10} = 0$.

\section{Note on collision depth}\label{appendix:collDepth}
\begin{figure*}
  \includegraphics[width=\linewidth]{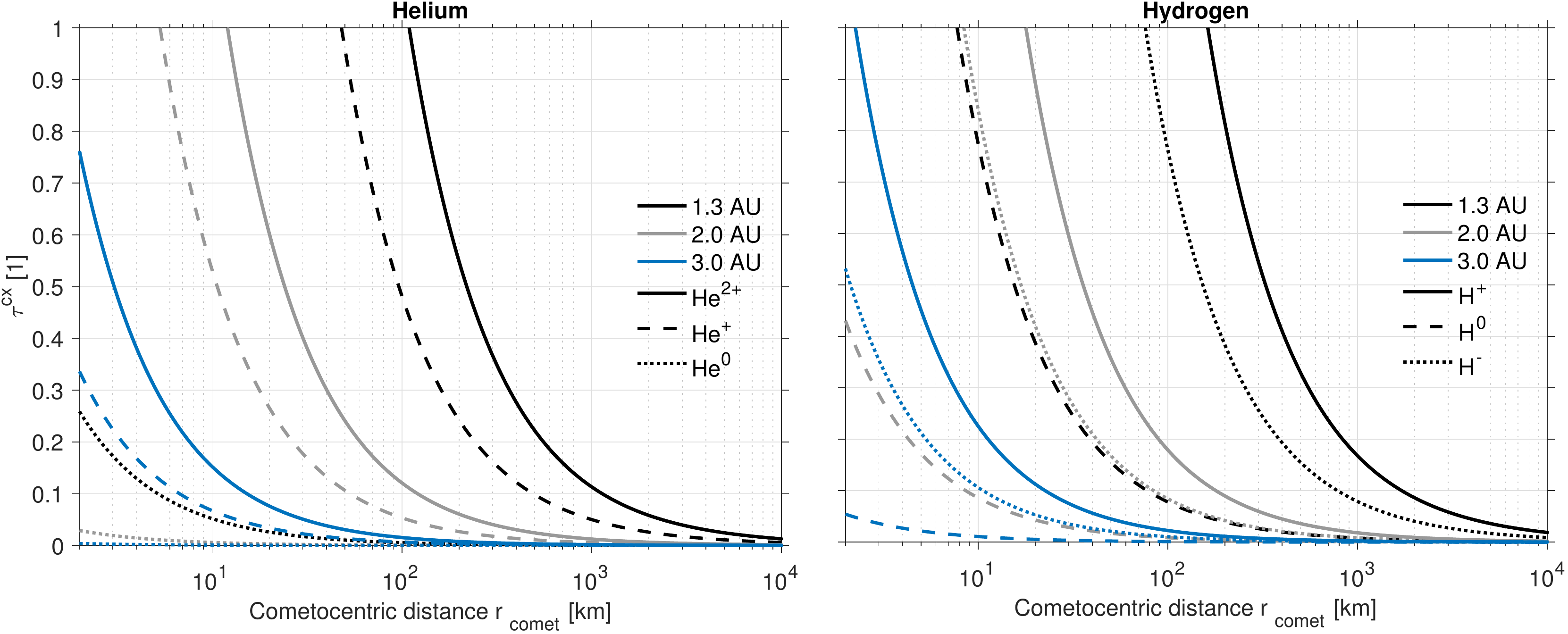}
     \caption{Charge-changing collision depth $\tau_i^\textnormal{cx}=\eta(r)\,\sigma_{i}$ for helium (left) and hydrogen species (right) in a comet 67P-like H$_2$O atmosphere for three typical heliocentric distances. The solar wind speed is assumed to be constant and equal to $400$~\kms{}.}
     \label{fig:tau_CX}
 \end{figure*}

Figure~\ref{fig:tau_CX} displays the charge-changing collision depth, defined as $\tau_i^\textnormal{cx}~=~\eta(r)\,\sigma_{i}$, where $\sigma_i$ is the sum of loss cross sections for each charge state $i$ (equation~[\ref{eq:sumXsections}]) of helium and hydrogen. In analogy with the Beer-Lambert optical depth, $\tau_\textnormal{cx} \geq 1$ represents the point where the atmosphere becomes effectively "opaque" to charge-changing reactions: particles experience significant charge-changing collisions. It depends on the projectile state, its energy, and on the neutral atmosphere, parameterized by a Haser-like model (see equation~[\ref{eq:HaserFull}], with $\varv_0=600$~m~s$^{-1}$). For a solar wind bulk speed of $400$~\kms{}, the atmosphere is almost transparent to He$^{0}$ and H$^0$ ENAs over the full range of cometocentric distances and for all heliocentric distances. This tendency is enhanced even further when decreasing the solar wind speed to $100$~\kms{}, with $\tau_i^\textnormal{cx}=1$ cometocentric distances decreasing by $25\%$ or more for each species (not shown). Comparatively, all positive ions will become efficiently charge-exchanged into lower charge states by the time they reach the typical cometocentric distances probed by the \emph{Rosetta} spacecraft.

\begin{acknowledgements}
      The work at University of Oslo was funded by the Norwegian Research Council "\emph{Rosetta}" grant No. 240000. Work at the Royal Belgian Institute for Space Aeronomy was supported by the Belgian Science Policy Office through the Solar-Terrestrial Centre of Excellence. Work at Ume\aa{} University was funded by SNSB grant 201/15. Work at Imperial College London was supported by STFC of UK under grant ST/K001051/1 and ST/N000692/1, ESA, under contract No.4000119035/16/ES/JD. The work at NASA/SSAI was supported by NASA Astrobiology Institute grant NNX15AE05G and by the NASA HIDEE Program.
C.S.W. would like to thank S. Barabash (IRF Kiruna, Sweden) for useful impetus on the work leading to the present study and for suggesting to investigate electron stripping processes at a comet. The authors thank the ISSI International Team "Plasma Environment of comet 67P after \emph{Rosetta}" for fruitful discussions and collaborations. C.S.W. thanks M.S.W. and L.S.W. for help in structuring this immense workload and for unwavering encouragements throughout these two years of work. 
Datasets of the \emph{Rosetta} mission can be freely accessed from ESA's Planetary Science Archive (\url{http://archives.esac.esa.int/psa}).
\end{acknowledgements}


\bibliographystyle{aa}
\bibliography{references} 

\end{document}